\documentclass[11pt]{article}
\usepackage[margin=1in]{geometry}
\usepackage{amsmath,amssymb,amsthm,mathtools,bm}
\usepackage{graphicx,booktabs,array,multirow,longtable}
\usepackage{natbib}
\usepackage{microtype}
\usepackage{enumitem}
\usepackage{hyperref}
\usepackage{caption}
\usepackage{float}
\usepackage{xcolor}
\usepackage{setspace}
\usepackage{mathrsfs}
\usepackage{bbm}
\hypersetup{colorlinks=true,citecolor=blue,linkcolor=blue,urlcolor=blue}
\newtheorem{theorem}{Theorem}
\newtheorem{proposition}{Proposition}
\newtheorem{corollary}{Corollary}
\newtheorem{lemma}{Lemma}
\theoremstyle{definition}
\newtheorem{definition}{Definition}
\newtheorem{assumption}{Assumption}
\theoremstyle{remark}
\newtheorem{remark}{Remark}
\newtheorem*{numericalresult}{Numerical Result 1}

\newcommand{\diag}{\operatorname{diag}}
\newcommand{\Cov}{\operatorname{Cov}}
\newcommand{\cum}{\operatorname{cum}}
\newcommand{\tr}{\operatorname{tr}}

\newcommand{\vech}{\operatorname{vech}}
\newcommand{\R}{\mathbb{R}}
\newcommand{\E}{\mathbb{E}}
\newcommand{\1}{\mathbf{1}}
\newcommand{\dd}{\,\mathrm{d}}
\newcommand{\T}{\mathsf{T}}

\title{\textbf{Common Signals, Spurious Cross-Excitation, and Identification\\
in Multivariate Hawkes Processes}}
\author{Kiseop Lee\thanks{Department of Statistics, Purdue University.}}
\author{Jiwon Jung\thanks{Department of Applied Mathematics, University of Washington, Seattle, WA, USA.}
\and
Kiseop Lee\thanks{Corresponding author. Department of Statistics, Purdue University, 150 N.\
University Street, West Lafayette, IN 47907, USA. Email: kiseop@purdue.edu.}}

\date{}

\begin{document}
\maketitle

\begin{abstract}
Multivariate Hawkes models are widely used to measure endogenous interactions in event data, but latent common signals can generate apparent excitation even without structural interaction. We characterize the resulting linear projection and construct stable, nonnegative Hawkes processes that share the mean intensity and Bartlett spectrum of common-signal models despite having different excitation mechanisms. This establishes a concrete form of second-order non-identification. We derive integrated third-order cumulants for Hawkes processes with Cox immigration and establish generic local identification of normalized structural parameters in two- and three-dimensional systems. An exact-equivalence Monte Carlo experiment demonstrates how third-order information can distinguish mechanisms with identical second-order structure, while revealing substantial finite-sample limitations under weak signals. For a restricted parametric family, we develop Whittle estimation from binned counts, establish consistency and asymptotic normality under appropriate identification conditions, and construct feasible sandwich standard errors accounting for estimated mean intensities. Simulations assess structural recovery and inference accuracy. An application to cryptocurrency transactions illustrates the sensitivity of inferred excitation to model specification and the limitations of structural interpretation based on spectral fit alone.
\end{abstract}


\section{Introduction}
When is a fitted Hawkes cross-excitation coefficient evidence of interaction between event streams, rather than their response to common information? This is an identification question, not merely a model-selection question. A predictor can assign nonzero excitation to a stream even when the data-generating process contains no event-to-event excitation, and successful fit need not resolve the ambiguity. The distinction matters especially in financial econometrics, where branching ratios and cross-kernels are often interpreted as reflexivity or directional influence.

A multivariate point process may exhibit clustering and directional dependence because events excite later events, because several streams respond to an unobserved environment, or because both mechanisms operate. These explanations need not be distinguished by the mean and covariance of the observed events. The distinction matters whenever a fitted excitation kernel is interpreted as a feature of the data-generating mechanism rather than solely as a prediction rule.

The underlying ambiguity is classical. \citet[Section~3]{Hawkes1971a} exhibited a self-exciting process with the same second-order properties as a doubly stochastic process. Building on that observation, we study how heterogeneous responses to one latent signal create a directional multivariate representation, when that representation is an admissible Hawkes kernel, and which additional information separates the resulting mechanisms.

We consider stationary multivariate Hawkes processes with Cox immigration. A nonnegative exogenous signal enters each baseline through a loading and a causal response measure; a separate kernel propagates excitation from observed events. The response measures may be densities or fixed delays. This separation allows us to compare the structural excitation kernel with the kernel of the best affine predictor based only on the observed event history.

\subsection{Related literature}
The spectral approach originates in \citet{Hawkes1971a,Hawkes1971b}, including the scalar equivalence example and multivariate Wiener--Hopf equations. The Poisson-cluster representation of \citet{HawkesOakes1974} supplies the branching interpretation. Spectral methods for shot-noise and random-excitation models are developed by \citet{BremaudMassoulie2002}. Within a correctly specified Hawkes class, \citet{BacryMuzy2016} characterize the causal kernel from first- and second-order statistics. There is no contradiction between that within-class uniqueness and equivalence between a Hawkes model and a model with latent immigration. Our canonical-factor formula uses this established prediction machinery; the additional calculation is the explicit delayed multivariate counterpart and the verification that all its kernel entries, its baseline, and its stability are admissible.

Stochastic baselines and hidden inputs are also established modeling devices. For example, \citet{MiscouridouEtAl2022} combine a log-Gaussian Cox background with Hawkes excitation, while \citet{WangShojaie2021,JinHuang2026} study latent confounding in point-process networks. Our model instead permits a general nonnegative exogenous factor and heterogeneous causal response measures. Exponential covariance is imposed only for the explicit filtering calculations, not for the integrated-cumulant formulas. The Gaussian system used in those calculations is a covariance surrogate, not the law of the nonnegative factor.

For higher-order information, \citet{AchabEtAl2018} use integrated second- and third-order cumulants to estimate the integrated kernel of a pure multivariate Hawkes process without specifying its temporal shape. Here we derive the additional terms due to Cox immigration and study the enlarged moment map containing both excitation and factor loadings. Exact rational witnesses inside a realizable parameter family establish generic local rank in dimensions two and three. A separate finite-window calculation for the symmetric Cox--Hawkes pair explains how higher-order information can distinguish two second-order-equivalent mechanisms; it is not a general test for the absence of excitation.

Our statistical results build on count-based Whittle estimation for ordinary Hawkes processes \citep{CheyssonLang2022}, multivariate mixing results \citep{BolyEtAl2023}, and multivariate spectral inference \citep{TangEtAl2026}. We provide the dependence argument for bounded Markov Cox immigration and retain the effect of estimating the mean inside the spectral model. The resulting sandwich variance contains both higher-order score covariances and score--mean cross-covariances. These limit results concern an explicitly restricted family and do not turn the cumulant identification theorem into a spectrum-only identification theorem.

Financial event data motivate the structural distinction. Branching and cross-excitation have been interpreted as reflexivity, contagion, and leadership \citep{FilimonovSornette2012,HardimanEtAl2013,AitSahaliaEtAl2015,JangEtAl2020,ChakravartyEtAl2021}. Misspecified background activity can distort univariate branching estimates \citep{FilimonovSornette2015}. More broadly, the separation of common responses from endogenous interaction parallels the distinction between correlated and endogenous effects in peer-effect models \citep{Manski1993,BramoulleEtAl2009}. These applications motivate the question, but the mathematical results concern point-process laws and their moment representations.

\subsection{Contributions and identification strategy}
First, we calculate the distortion of the observed-history affine predictor. With exponential factor covariance and exponential response measures, the projected kernel is obtained from a finite-dimensional linear filter, while the structural excitation kernel need not have a parametric form. For two equal-rate streams responding at different fixed delays and with no structural excitation, we give all four projected kernels explicitly. They are nonnegative and stable for every positive signal strength and delay in the stated family. Their integrated directional asymmetry increases with signal strength. This admissible delayed construction, rather than spectral factorization or scalar second-order ambiguity alone, is the main representation result.

Second, we characterize information beyond that representation. The integrated cumulant formulas yield generic local identification of the normalized branching matrix, factor loadings, and long-run third factor cumulant for two and three streams. They do not identify kernel shapes or the full latent law. For a symmetric two-state factor, we also show exact finite-window third-order separation from its Hawkes counterpart and evaluate that distinction in a controlled experiment. Observing an exogenous proxy provides a different route: the linear-filter error covariance decreases, and the residual kernel distortion vanishes in an ideal high-precision limit, without an entrywise monotonicity claim.

The identifying information differs across our results. The exact counterpart rules out structural interpretation from unrestricted first- and second-order features alone. Integrated third-order cumulants add information and can identify normalized parameters locally; in contrast, frequency-domain estimation is feasible only after imposing a narrower parametric structure and separate spectral-identification conditions. The latter is not a plug-in estimator for the former identification theorem.

Third, for a bounded Markov-factor family with exponential excitation and bounded delays, we establish Whittle consistency to the spectral identified set and asymptotic normality for consistent regular interior fits. A feasible long-run covariance estimator includes the estimated-mean correction. Simulations distinguish three questions: separation of fixed second-order-equivalent mechanisms, recovery under parametric restrictions, and calibration of interior standard errors. The equivalent-pair experiment is the central numerical illustration; its weak-signal results and the finite-sample undercoverage of delay intervals are reported alongside the favorable results.

A short SOL spot--perpetual illustration examines what a spectral calibration actually uses. Matched-basis profiles and unfitted hourly moments show why agreement on within-hour frequencies need not validate long-run attribution. It is not offered as empirical proof of exact observational equivalence. The extensive timing, profile, covariance-adequacy, and proxy diagnostics are collected in the Online Supplement.

Section~\ref{sec:model} specifies the model. Sections~\ref{sec:spurious} and \ref{sec:nonid} develop the projection and its admissible counterparts; Section~\ref{sec:identification} treats additional identifying information. Section~\ref{sec:estimation} gives restricted estimation and inference, Section~\ref{sec:simulation} presents the numerical experiments, and Section~\ref{sec:empirical} gives the data illustration. For transparency, the projection argument remains beside Theorem~\ref{thm:closed}; the two central mathematical verification arguments are retained in Appendices~\ref{app:delayMain} and \ref{app:certificatesMain}. The finite-window separation is proved in Section~\ref{sec:equivalentPairMC}; Appendix~\ref{app:certificatesMain} gives the identification certificates, and the remaining proofs are in the Online Supplement.

\section{Model}\label{sec:model}
Let $N=(N^1,\ldots,N^d)$, where $N^i((a,b])$ counts events of type $i$ between times $a$ and $b$. We work with a stationary event process; stationarity refers to the distribution of the event configuration under time shifts, not to the cumulative counts from a fixed origin. Conditional on the latent environment, its intensity is
\begin{equation}\label{eq:model}
\lambda_t^i
=\mu_i
+\beta_i\int_{[0,\infty)}Y_{t-u}\,g_i(\dd u)
+\sum_{j=1}^d\int_{(-\infty,t)}\varphi_{ij}(t-s)\,\dd N_s^j.
\end{equation}
Here $Y$ is a latent common signal, $g_i$ is a probability measure on $[0,\infty)$ describing the timing and shape of type $i$'s response to that signal, $\beta_i\ge0$ scales the magnitude of the response, and $\varphi_{ij}\ge0$ are excitation kernels. The probability-measure formulation accommodates both distributed
responses, $g_i(du)=g_i(u)\,du$, and fixed response delays: a delay
$\delta_i\ge0$ is represented by $g_i=\delta_{\delta_i}$. Write
\[
\Phi_{ij}=\int_0^\infty \varphi_{ij}(u)\,\dd u,
\qquad
\widehat f(\omega)=\int e^{-i\omega u}f(\dd u)
\]
for integrated kernels and Fourier transforms.

\begin{assumption}[Basic model]\label{ass:basic}
The following conditions hold.
\begin{enumerate}
\item[(i)] The latent signal $Y$ is stationary, nonnegative, exogenous to $N$, and has finite second moment.

\item[(ii)] Conditional on the path of $Y$, $N$ is a multivariate Hawkes process with time-varying baseline
\[
\nu_t^i
=\mu_i+\beta_i\int_{[0,\infty)}Y_{t-u}\,g_i(\dd u),
\qquad \mu_i,\beta_i\ge0.
\]

\item[(iii)] The excitation kernels $\varphi_{ij}$ are nonnegative and satisfy
\[
\varphi_{ij}\in L^1(\R_+)\cap L^2(\R_+),
\qquad \rho(\Phi)<1,
\]
and the mean immigration rates $\bar\nu_i=\mu_i+\beta_i\E Y_0$ are strictly positive.
\end{enumerate}
\end{assumption}

The conditional specification is understood in the Cox-immigration construction: the full signal path is fixed before the baseline events and their independent offspring clusters are generated. Thus exogeneity does not mean that $Y$ and $N$ are independent. Assumption~\ref{ass:basic} separates the two sources of dependence in \eqref{eq:model}: the common environment enters through the random baselines $\nu_t^i$, while dependence propagated by observed events enters through the Hawkes kernels. The intensity in \eqref{eq:model} is conditional on the latent environment; its best linear projection on the observed history need not equal the conditional intensity with respect to the observed filtration. The stationary mean intensity is
\[
\Lambda=(I-\Phi)^{-1}\bar\nu,
\qquad
\bar\nu_i=\mu_i+\beta_i\E Y_0,
\]
where we use that each response measure $g_i$ has unit mass.

The closed-form second-order results in Sections~\ref{sec:spurious}--\ref{sec:nonid} additionally use the exponential covariance specification
\begin{equation}\label{eq:expcov}
\Cov(Y_t,Y_{t+\tau})=v e^{-\kappa|\tau|},\qquad v,\kappa>0.
\end{equation}
This restriction is not part of the basic model. It is satisfied, for example, by a CIR process and by a nonnegative two-state Markov chain. A Gaussian OU process has the same covariance structure and is useful as a second-order surrogate, but a nondegenerate Gaussian OU process is not itself a nonnegative Cox baseline.

\begin{definition}[Information and excitation]
The common-information channel is described by $(\beta_i,g_i)$. The response measure $g_i$ determines the timing and shape of type $i$'s response to the latent signal $Y$, while $\beta_i$ determines its magnitude. The excitation channel is described by $\varphi_{ij}$: the kernel $\varphi_{ij}$ measures how past type-$j$ events affect the future intensity of type $i$. Off-diagonal kernels describe cross-excitation; diagonal kernels describe self-excitation.
\end{definition}
When $i\ne j$, cross-excitation may be interpreted economically as imitation or contagion, depending on the application. Agents who run similar models trained on similar data can instead move together because they respond to the same signal; that dependence belongs to the common-information channel even if it resembles herding. If type $i$ reacts to a price move caused by type $j$'s trades, the response propagates through $N^j$ and is classified as price-mediated cross-excitation. Exogeneity of $Y$ remains a maintained modeling restriction; this classification does not establish it from observed order flow.

\begin{lemma}[Bartlett spectrum]\label{lem:spectrum}
Under Assumption~\ref{ass:basic}, if $Y$ has spectral density $S_Y$, then the Bartlett spectral density matrix of $N$ is
\begin{equation}\label{eq:spectrum}
S(\omega)=\bigl(I-\widehat\varphi(\omega)\bigr)^{-1}
\left[D_\Lambda+b(\omega)S_Y(\omega)b(\omega)^*\right]
\left[\bigl(I-\widehat\varphi(\omega)\bigr)^{-1}\right]^*,
\end{equation}
where $D_\Lambda=\diag(\Lambda)$, $b_i(\omega)=\beta_i\widehat g_i(\omega)$, and $^*$ denotes conjugate transpose. Under \eqref{eq:expcov}, $S_Y(\omega)=2v\kappa/(\kappa^2+\omega^2)$.
\end{lemma}
A pure delay $g_i=\delta_{\delta_i}$ gives $b_i(\omega)=\beta_i e^{-i\omega\delta_i}$, and an exponential response $g_i(\dd u)=\kappa_i e^{-\kappa_i u}\dd u$, with $\kappa_i>0$, gives $b_i(\omega)=\beta_i\kappa_i/(\kappa_i+i\omega)$. In both cases heterogeneous response speed enters the cross-spectrum through phase.

\section{Spurious cross-excitation}\label{sec:spurious}
The question here is how the omitted signal changes a predictor based only on observed events. The intensity $\lambda_t$ in \eqref{eq:model} generally cannot be recovered from the history of $N$ alone. We restrict attention to rules that are affine in past counts, as in the Hawkes specification. This restriction lets us characterize their coefficients from the mean and covariance of $N$, without specifying its full distribution.

\begin{definition}[Hawkes-form linear projection and projected kernel]\label{def:projection}
Let $\mathcal H_{t-}$ be the closed linear span in $L^2(\Omega,\mathcal F,\mathbb P)$ of the centered past counts
\[
N^j((a,b])-\Lambda_j(b-a),\qquad a<b<t,\quad j=1,\ldots,d.
\]
Writing $\Pi_{t-}$ for the componentwise orthogonal projection onto this space, define
\[
\widetilde\lambda_t=\Lambda+\Pi_{t-}(\lambda_t-\Lambda).
\]
Thus $\widetilde\lambda_t$ minimizes $\E\|\lambda_t-q_t\|^2$ over predictors whose coordinates are constants plus elements of $\mathcal H_{t-}$. When it admits the representation
\[
\widetilde\lambda_t
=\widetilde\mu+\int_{(-\infty,t)}\widetilde\varphi(t-s)\,\dd N_s,
\qquad
\widetilde\mu=(I-\widetilde\Phi)\Lambda,
\quad
\widetilde\Phi=\int_0^\infty\widetilde\varphi(u)\,\dd u,
\]
with a deterministic causal, integrable, possibly signed matrix kernel $\widetilde\varphi$, we call $\widetilde\lambda$ the \emph{Hawkes-form linear projection} and $\widetilde\varphi$ its \emph{projected kernel}.
\end{definition}
Here ``linear'' refers to past counts, not time or model parameters, and ``best'' refers to mean-square error. The definition projects the square-integrable rate $\lambda_t$, rather than treating $\dd N_t/\dd t$ as an ordinary random variable. The compensator identity links this rate projection to prediction of event increments and to a linear least-squares contrast; the details are given in Online Supplement Section S.1.2 \citep{HansenEtAl2015}.

The mean and covariance of $N$ determine the projection coefficients; evaluating the predictor also uses the realized event history. This is a rule defined by the probability law, not an estimator fitted to a finite sample. In general it differs both from the latent-environment intensity and from the full conditional intensity given the observed past. Its kernel need not be nonnegative, and a constrained Hawkes maximum-likelihood fit need not have the same target.

The following standard innovation-factor relation connects prediction to the spectrum in Lemma~\ref{lem:spectrum} \citep{Hawkes1971b,BacryMuzy2016}. It does not assert that the resulting kernel is nonnegative. That additional requirement is checked explicitly for the delayed counterpart below.

\begin{lemma}[Canonical factor representation]\label{lem:factor}
Suppose $S(\omega)=L(\omega)D_\Lambda L(\omega)^*$ is a normalized minimum-phase factorization with $L(\omega)=I+\widehat\ell(\omega)$ and $L(\omega)^{-1}=I+\widehat h(\omega)$, where $\ell,h$ are causal matrix kernels in $L^1(\mathbb R_+)\cap L^2(\mathbb R_+)$. Then
\begin{equation}\label{eq:projectionfactor}
\widehat{\widetilde\varphi}(\omega)=I-L(\omega)^{-1}.
\end{equation}
\end{lemma}

Lemma~\ref{lem:factor} is general but implicit: it identifies the projected kernel through a spectral factor. To see the distortion created by the latent signal more transparently, we now specialize only the common-information response to an exponential form. This yields a finite-dimensional filtering problem and an explicit formula for the difference between structural and projected excitation.

\subsection{Closed form with exponential responses}\label{sec:closed}
To make the projected kernel explicit, suppose the latent signal has the exponential covariance in \eqref{eq:expcov} and type $i$ has an exponential response measure,
$g_i(\dd u)=\kappa_i e^{-\kappa_i u}\dd u$ for $u\ge0$. Thus the common-information component of its intensity is a weighted average of past signal values: more recent values receive greater weight, and the weights decay at rate $\kappa_i$. The response timescale is $1/\kappa_i$, so larger $\kappa_i$ means faster adjustment. This specification is useful because these weighted averages obey finite-dimensional linear dynamics, allowing us to compute the Hawkes-form projection by linear filtering rather than solving an infinite-history prediction problem. For this subsection only, write
\[
X_t^i:=\int_0^\infty \kappa_i e^{-\kappa_i u}Y_{t-u}\,\dd u
\]
for the exponentially filtered signal seen by type $i$, and set
\[
z=(Y-\E Y,X^1-\E Y,\ldots,X^d-\E Y)^\T\in\R^{d+1}.
\]
For linear prediction, only the mean and covariance of $z$ are needed. They admit the following state-space representation:
\begin{equation}\label{eq:state}
\dd z_t=A z_t\dd t+\dd W_t,
\qquad
A=\begin{pmatrix}-\kappa&0\\ \bm\kappa&-\diag(\bm\kappa)\end{pmatrix},
\qquad
\E(\dd W_t\dd W_t^\T)=Q\dd t,
\end{equation}
where $\bm\kappa=(\kappa_1,\ldots,\kappa_d)^\T$, $Q=\diag(2\kappa v,0,\ldots,0)$, and $W$ is a zero-mean orthogonal-increment input with the displayed covariance rate. A Gaussian surrogate realizes this covariance representation; the actual factor $Y$ need not be Gaussian. Let
\[
H=[\,0\ \ \diag(\beta)\,]\in\R^{d\times(d+1)}.
\]
Let $P$ be the stabilizing solution of
\begin{equation}\label{eq:riccati}
AP+PA^\T+Q-PH^\T D_\Lambda^{-1}HP=0,
\end{equation}
and define
\[
K=PH^\T D_\Lambda^{-1},\qquad F=A-KH.
\]
The matrix $H$ maps latent states to the centered common-information component of intensity. To isolate that component, first remove the structural excitation from the centered event observations. Their remaining observation-noise covariance rate is $D_\Lambda$; $P$ is the steady-state linear-filter error covariance, $K$ is its gain, and $F$ governs the filter dynamics. This is a calculation at the specified model parameters, not an estimation algorithm that assumes $\varphi$ is observed. Re-expressing the filter in terms of the original event history gives the projected kernel in the next theorem.

\begin{theorem}[Closed-form projection under exponential responses]\label{thm:closed}
Under Assumption~\ref{ass:basic}, \eqref{eq:expcov}, and exponential response measures, define
\[
\Psi(u)=He^{Fu}K,\qquad u>0.
\]
Then, for every $\omega\in\R$,
\begin{equation}\label{eq:multbias}
I-\widehat{\widetilde\varphi}(\omega)
=\bigl(I-\widehat\Psi(\omega)\bigr)
 \bigl(I-\widehat\varphi(\omega)\bigr),
\qquad
\widehat\Psi(\omega)=H(i\omega I-F)^{-1}K.
\end{equation}
Equivalently,
\begin{equation}\label{eq:kernelbias}
\widetilde\varphi=\varphi+\Psi-\Psi*\varphi,
\qquad
\widetilde\Phi=\Phi+\bar\Psi(I-\Phi),
\qquad
\bar\Psi=-HF^{-1}K.
\end{equation}
No parametric shape is required for the structural excitation kernel $\varphi$ beyond the regularity in Assumption~\ref{ass:basic}.
\end{theorem}

\begin{proof}
For centered increments, the state representation gives
\[
\dd\widetilde N_t=\{Hz_t+(\varphi*\dd\widetilde N)_t\}\dd t+\dd M_t.
\]
The causal resolvent $\mathcal R=\sum_{n\ge1}\varphi^{*n}$ is integrable because the kernels are nonnegative and $\rho(\Phi)<1$. It is also square integrable under Assumption~\ref{ass:basic}. Writing $\delta_0I$ for the identity convolution kernel, we have that $\delta_0I-\varphi$ has the causal stable inverse $\delta_0I+\mathcal R$: residualization preserves the closed linear span of past increments. Residualize the observations:
\[
\dd O_t=\dd\widetilde N_t-(\varphi*\dd\widetilde N)_t\dd t=Hz_t\dd t+\dd M_t.
\]
Since $A$ is Hurwitz and $D_\Lambda$ is positive definite, the filtering Riccati equation has a unique positive-semidefinite stabilizing solution. The second-order state/observation covariance is identical to that of a Gaussian linear system with state matrix $A$, process covariance $Q$, observation matrix $H$, and observation-noise covariance rate $D_\Lambda$. Linear projections depend only on these covariances, so the stationary optimal linear filter \citep{AndersonMoore1979} has covariance $P$ solving \eqref{eq:riccati} and
\[
\widehat z_t=\int_{-\infty}^t e^{F(t-s)}K\,\dd O_s.
\]
The affine $L^2$ projection of the rate is therefore
\[
\Lambda+H\widehat z_t+(\varphi*\dd\widetilde N)_t
=\Lambda+\{\Psi*(\delta_0I-\varphi)*\dd\widetilde N\}_t+(\varphi*\dd\widetilde N)_t.
\]
Hence the kernel acting on $\dd\widetilde N$ is $\varphi+\Psi-\Psi*\varphi$, which gives \eqref{eq:multbias}--\eqref{eq:kernelbias}. This is the Hawkes-form linear-projection calculation from Definition~\ref{def:projection}; Gaussianity is not imposed on the actual point process or on the latent Cox factor.
\end{proof}
The decomposition in \eqref{eq:kernelbias} separates the two sources of apparent excitation. The structural kernel $\varphi$ is present even if the common signal were observed, while $\Psi-\Psi*\varphi$ is the additional term created by predicting the omitted common component from past events. Thus $\Psi$ is a bias-generating filter, not in general the bias itself. When $\varphi\equiv0$, the distinction disappears and the entire projected kernel is generated by the common signal.

Theorem~\ref{thm:closed} is a linear-prediction theorem. It does not assert that each entry of $\widetilde\varphi-\varphi$ is nonnegative, that $\rho(\widetilde\Phi)\ge \rho(\Phi)$, or that a constrained Hawkes MLE has $\widetilde\varphi$ as its exact population target. The nonnegative examples below show when this predictive representation also defines a genuine Hawkes counterpart.

The univariate case makes the size of this effect particularly transparent. It shows that even with no structural excitation, persistent common variation alone can produce an apparently highly endogenous Hawkes fit.

\begin{corollary}[One dimension]\label{cor:oned}
Under Assumption~\ref{ass:basic} and \eqref{eq:expcov}, let $d=1$ and let the response be instantaneous. Write $n=\int_0^\infty\varphi(u)\,\dd u$ and $\widetilde n=\int_0^\infty\widetilde\varphi(u)\,\dd u$. Define
\[
F_c=1+\frac{2\beta^2v}{\kappa\Lambda},
\qquad
\gamma=\kappa\sqrt{F_c}.
\]
Then
\[
\Psi(u)=(\gamma-\kappa)e^{-\gamma u},
\qquad
\widetilde n=1-\frac{1-n}{\sqrt{F_c}}.
\]
With $n=0$, the second-order counterpart is a genuine Hawkes process with exponential kernel, branching ratio $1-F_c^{-1/2}$, and decay rate $\kappa\sqrt{F_c}$.
\end{corollary}
The scalar counterpart is consistent with the classical exponential second-order ambiguity in \citet[Section~3]{Hawkes1971a}. As the regimes become persistent ($\kappa\to0$ with $v$ fixed), $F_c\to\infty$ and $\widetilde n\to1$, reproducing the apparent criticality documented by \citet{FilimonovSornette2015}. With no excitation, mean intensity one event per second, standard deviation $0.5$ for the signal-driven intensity variation, and covariance persistence time ten minutes, $F_c=301$ and $\widetilde n\approx0.94$; the spurious kernel decays over about $35$ seconds.

\subsection{Direction: weak signal and an exact pure-delay benchmark}
The univariate calculation explains apparent persistence, but our main multivariate concern is direction. If two streams respond to the same signal at different speeds, does the faster stream appear to excite the slower one? We answer this first in a weak-signal expansion, which reveals the mechanism for general response measures, and then in a pure-delay benchmark where the directional effect can be computed exactly at any signal strength.

Scale the common-signal loadings as $\beta_\varepsilon=\varepsilon b$, with $b$ fixed, and let $\Gamma_\varepsilon(u)$ denote the covariance matrix of the resulting common-information components. Thus
\[
[\Gamma_\varepsilon(u)]_{ij}
=\varepsilon^2 b_i b_j\,\Cov\!\left(
\int Y_{t-v}\,g_i(\dd v),
\int Y_{t-u-w}\,g_j(\dd w)
\right).
\]
Let $D_\varepsilon=D_{\Lambda_\varepsilon}$. For general response measures, $\Psi_\varepsilon$ denotes the causal linear-prediction kernel of the residual observations after structural excitation is removed; their spectrum is $D_\varepsilon+\widehat\Gamma_\varepsilon$. This extends the residual filter used in Theorem~\ref{thm:closed} beyond exponential responses.

\begin{proposition}[Weak signal]\label{prop:weak}
Assume $\Gamma_\varepsilon(u)=\varepsilon^2\Gamma_0(u)$ with $\Gamma_0\in L^1(\R)\cap L^\infty(\R)$, and that $D_\varepsilon^{-1}$ is uniformly bounded for small $\varepsilon$. Then the causal residual-factor projected kernel satisfies
\begin{equation}\label{eq:weak}
\left\|\Psi_\varepsilon-\Gamma_\varepsilon(\cdot)D_\varepsilon^{-1}\right\|_{L^1(\R_+)}
+
\left\|\Psi_\varepsilon-\Gamma_\varepsilon(\cdot)D_\varepsilon^{-1}\right\|_{L^\infty(\R_+)}
=O(\varepsilon^4).
\end{equation}
\end{proposition}

To leading order in signal strength, Proposition~\ref{prop:weak} says that the residual filter $\Psi_\varepsilon$ inherits the positive-lag covariance of the omitted common-information component, scaled by event intensity. When structural excitation is absent, it is the entire projected kernel. Direction therefore comes from an asymmetry in that covariance across positive lags. Pure delays make this especially easy to see.

\begin{corollary}[Weak-signal directional asymmetry]\label{cor:weakdir}
Under Assumption~\ref{ass:basic} and \eqref{eq:expcov}, let $d=2$, $\beta_1=\beta_2=\beta>0$, $\Lambda_1=\Lambda_2=\Lambda$, and $g_1=\delta_0$, $g_2=\delta_\Delta$ with $\Delta>0$. Let
\[
r=\frac{2\beta^2v}{\kappa\Lambda},\qquad x=\kappa\Delta.
\]
Then
\[
\bar\Psi_{21}=\frac r2(2-e^{-x})+O(r^2),
\qquad
\bar\Psi_{12}=\frac r2 e^{-x}+O(r^2),
\]
and
\begin{equation}\label{eq:weakdir}
\frac{\bar\Psi_{21}-\bar\Psi_{12}}{\bar\Psi_{21}+\bar\Psi_{12}}
=1-e^{-x}+O(r).
\end{equation}
The ratio in \eqref{eq:weakdir} is a directional asymmetry index; the fraction of cross-mass pointing fast-to-slow is one half of one plus this index.
\end{corollary}

The weak-signal result already shows that a reaction-time gap creates fast-to-slow apparent excitation. The next result strengthens that statement in the symmetric pure-delay benchmark: it removes the weak-signal approximation, verifies that the projected kernel is genuinely nonnegative and stable, and therefore produces a second-order-equivalent Hawkes process rather than merely a signed linear predictor.

\begin{corollary}[Exact symmetric pure-delay case]\label{cor:exactdelay}
In the setting of Corollary~\ref{cor:weakdir}, assume in addition that $\varphi\equiv0$. Put
\[
a=\kappa\sqrt{1+2r},\qquad b=\kappa\sqrt{1+r},\qquad q=\frac{a-\kappa}{2},\qquad c=\frac{a+\kappa}{2},
\]
\[
D=\cosh(b\Delta)+\frac cb\sinh(b\Delta),\qquad
E(t)=\cosh(bt)+\frac\kappa b\sinh(bt),\qquad
F_0(t)=\cosh(bt)+\frac ab\sinh(bt).
\]
The projected kernel, which here equals $\widetilde\varphi=\Psi$, is
\begin{align}
\Psi_{11}(u)&=\frac qD
\begin{cases}
F_0(\Delta-u),&0<u<\Delta,\\
e^{-a(u-\Delta)},&u\ge\Delta,
\end{cases}
\label{eq:delay11}\\
\Psi_{12}(u)&=\frac qD e^{-au},\label{eq:delay12}\\
\Psi_{21}(u)&=\frac qD
\begin{cases}
E(u),&0<u<\Delta,\\
E(\Delta)e^{-a(u-\Delta)},&u\ge\Delta,
\end{cases}
\label{eq:delay21}\\
\Psi_{22}(u)&=\frac{qE(\Delta)}D e^{-au}.\label{eq:delay22}
\end{align}
All four entries are nonnegative and the integrated matrix $\bar\Psi$ has row sums strictly below one. Hence there exists a genuine stationary Hawkes process with kernel $\Psi$ and positive baseline $\Lambda(I-\bar\Psi)\1$ that has exactly the same mean and Bartlett spectrum as the no-excitation common-factor process.

Moreover, with $x=\kappa\Delta$, $\alpha=\sqrt{1+2r}$, and $b_0=\sqrt{1+r}$, define
\begin{equation}\label{eq:Qratio}
Q(r,x)=1+\left(1+\frac{\alpha}{b_0^2}\right)\{\cosh(b_0x)-1\}
+\frac{\alpha+1}{b_0}\sinh(b_0x).
\end{equation}
Then
\[
\frac{\bar\Psi_{21}}{\bar\Psi_{12}}=Q(r,x),
\qquad
\mathcal D(r,x):=
\frac{\bar\Psi_{21}-\bar\Psi_{12}}{\bar\Psi_{21}+\bar\Psi_{12}}
=\frac{Q(r,x)-1}{Q(r,x)+1}.
\]
For every $x>0$, $\mathcal D(r,x)$ is strictly increasing in $r>0$, with
\[
\lim_{r\downarrow0}\mathcal D(r,x)=1-e^{-x},
\qquad
\lim_{r\to\infty}\mathcal D(r,x)=1.
\]
At $x=0$ the asymmetry is identically zero for $r>0$. At $r=0$ both cross-masses vanish, so the directional ratio is understood only by its limit.
\end{corollary}

The full proof is in Appendix~\ref{app:delayMain}. It separates the causal filtering calculation from the two admissibility checks: positivity of every kernel entry and strict positivity of both row-sum defects. These checks hold throughout the stated symmetric family, not merely at numerically selected parameter values. Equal loadings, equal mean rates, and two fixed response delays are substantive restrictions; no corresponding positivity assertion is made for arbitrary multivariate responses.

\begin{figure}[t]
\centering
\includegraphics[width=0.48\textwidth]{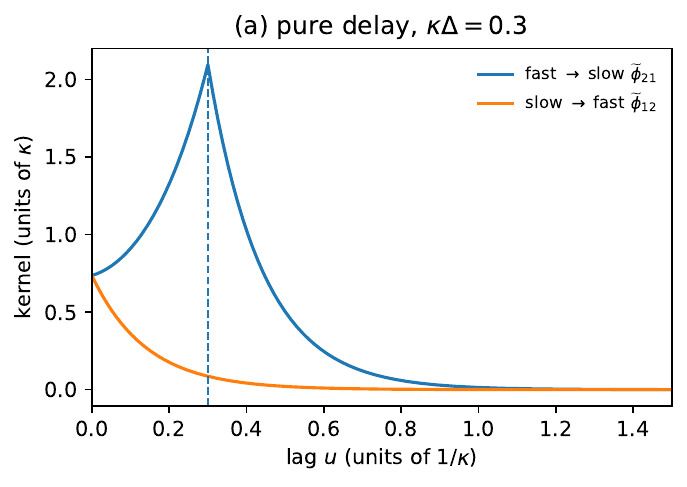}\hfill
\includegraphics[width=0.48\textwidth]{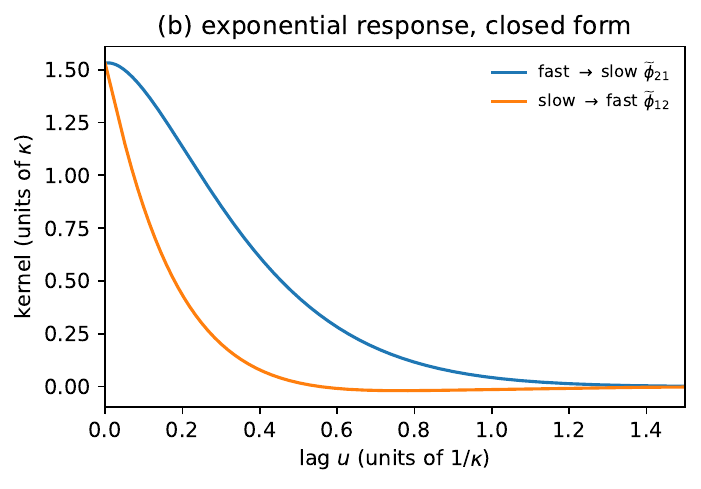}
\caption{Spurious cross-excitation kernels when there is no imitation ($\varphi\equiv0$). Two agents have equal loadings and mean intensities $\Lambda=50\kappa$, with latent covariance $ve^{-\kappa|\tau|}$ and $r=2\beta^2v/(\kappa\Lambda)=25$. (a) Pure delays $0$ and $\Delta=0.3/\kappa$, evaluated from the normalized kernels in Corollary~\ref{cor:exactdelay}. (b) Exponential responses with $\kappa_1=50\kappa$ and $\kappa_2=\kappa/0.3$, from Theorem~\ref{thm:closed}. In panel (b) the eigenvalues of $F$ are approximately $-49.7$ and $-4.6\pm1.6i$ in units of $\kappa$. Panel (b) is an unconstrained linear projection and has small signed tails; the nonnegative Hawkes counterpart is guaranteed for panel (a).}
\label{fig:kernels}
\end{figure}

\begin{figure}[t]
\centering
\includegraphics[width=0.58\textwidth]{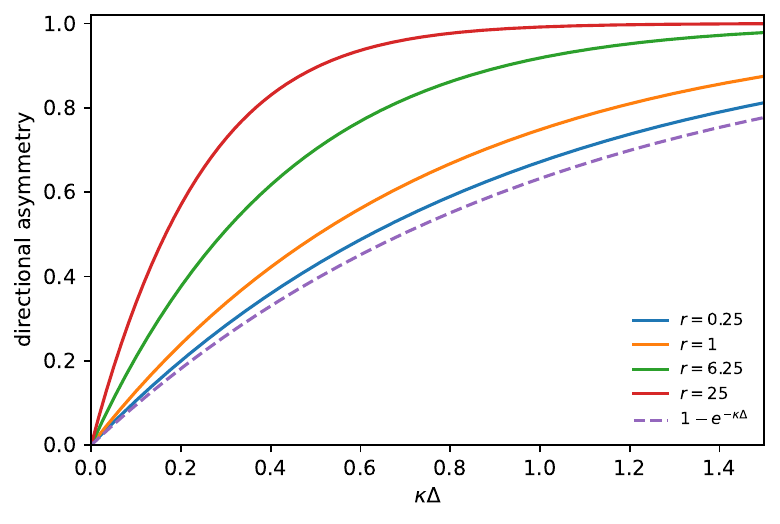}
\caption{Exact directional asymmetry $(\bar\Psi_{21}-\bar\Psi_{12})/(\bar\Psi_{21}+\bar\Psi_{12})$ in the symmetric pure-delay model against $\kappa\Delta$, for signal-to-noise ratios $r=2\beta^2v/(\kappa\Lambda)$. Dashed: the weak-signal limit $1-e^{-\kappa\Delta}$.}
\label{fig:direction}
\end{figure}

Figure~\ref{fig:kernels} shows the shape of the apparent cross-excitation produced by heterogeneous responses to a common signal. Both panels set the structural excitation kernel to zero, $\varphi\equiv0$, so every nonzero projected cross-kernel is due to the omitted signal. In panel (a), type 1 responds immediately, while type 2 responds after a fixed delay $\Delta$. Past events of the fast type reveal changes in the signal before the corresponding slow response occurs. The projected fast-to-slow kernel $\widetilde\varphi_{21}=\Psi_{21}$ consequently rises over the interval $(0,\Delta)$ and decays after $\Delta$; the reverse kernel $\widetilde\varphi_{12}=\Psi_{12}$ decays from lag zero. Panel (b) replaces fixed delays by exponential response weights. The agents now react gradually, yet their different response timescales still generate a pronounced asymmetry. The hump in the fast-to-slow kernel reflects the dynamics of the causal linear filter in Theorem~\ref{thm:closed}, rather than any direct excitation between agents. Thus neither the direction nor the temporal shape of a projected cross-kernel can, by itself, be interpreted as structural influence.

Figure~\ref{fig:direction} quantifies this asymmetry in the symmetric pure-delay benchmark. Its horizontal axis $x=\kappa\Delta$ measures the reaction-time gap relative to the persistence timescale $1/\kappa$ of the common signal; $r$ measures the strength of common-signal variation relative to the event-noise level. With simultaneous responses ($x=0$), the two projected cross-masses are equal and the directional asymmetry index $\mathcal D(r,x)$ is zero. As the response gap grows, events of the fast type become more informative about the slow type's subsequent activity, increasing the fast-to-slow asymmetry. The dashed curve $1-e^{-x}$ is the weak-signal limit from Corollary~\ref{cor:weakdir}; the remaining curves illustrate the exact finite-signal expression from Corollary~\ref{cor:exactdelay}. The latter also proves that, for each $x>0$, $\mathcal D(r,x)$ increases with $r$ and tends to one as $r$ grows. An index near one means that nearly all of the \emph{projected cross-mass} points fast-to-slow, not that nearly all events are caused by the fast type. Throughout this example, structural cross-excitation remains identically zero.

\subsection{Apparent criticality in several dimensions}
For a multivariate Hawkes process, criticality refers to the stability boundary $\rho(\Phi)=1$, where $\Phi$ is the integrated structural excitation matrix. We ask whether omitting the common signal can make the projected branching matrix $\widetilde\Phi$ appear close to this boundary even when $\Phi$ is not. When the projected kernel is nonnegative and defines a stable Hawkes counterpart, its spectral radius summarizes branching amplification. A signed projected kernel does not have that offspring interpretation.

Taking determinants in \eqref{eq:multbias} at $\omega=0$ gives
\[
\det(I-\widetilde\Phi)=\det(I-\bar\Psi)\det(I-\Phi).
\]
This identity factors the zero-frequency transfer determinant. It is not a general distance-to-criticality formula and does not by itself imply monotonicity of $\rho(\widetilde\Phi)$. For the nonnegative pure-delay counterpart in Figure~\ref{fig:kernels}(a), $\rho(\widetilde\Phi)\approx0.832$ although $\Phi=0$. Here the apparent amplification has a genuine Hawkes interpretation, but the data-generating model has no structural excitation. An endogeneity interpretation therefore requires information about the common-factor component, not just a fitted branching matrix.

\section{Second-order representation and its identification limits}\label{sec:nonid}
The preceding formulas describe observed-history predictors. To interpret such a representation as another point process, its kernel and baseline must also satisfy the Hawkes positivity and stability restrictions. The next theorem records this representation-to-process step using standard spectral and cluster theory \citep{Hawkes1971b,HawkesOakes1974}. The explicit verification for heterogeneous delays is supplied by Corollary~\ref{cor:exactdelay}.

Here second-order equivalence means equality of the mean vectors and covariance measures, or equivalently the Bartlett spectra when spectral densities exist. It does not mean equality of the full laws. The scalar precedent is \citet[Section~3]{Hawkes1971a}; our use of the criterion is to exhibit an admissible directional multivariate family.

\begin{theorem}[Second-order representation and Hawkes counterpart]\label{thm:nonid}
For a mean vector $\Lambda>0$, suppose its spectrum $S$ admits a normalized minimum-phase factorization
\[
S(\omega)=L(\omega)D_\Lambda L(\omega)^*,
\qquad
L(\omega)=I+\widehat\ell(\omega),
\qquad
L(\omega)^{-1}=I+\widehat h(\omega),
\]
where $\ell$ and $h$ are causal matrix kernels in $L^1(\mathbb R_+)\cap L^2(\mathbb R_+)$. Define
\[
\widehat{\widetilde\varphi}(\omega)=I-L(\omega)^{-1}.
\]
Then $\widetilde\varphi$ is the projected kernel of the normalized causal linear predictor, unique up to equality almost everywhere, and
\begin{equation}\label{eq:secondorderrep}
S(\omega)=\bigl(I-\widehat{\widetilde\varphi}(\omega)\bigr)^{-1}
D_\Lambda
\left[\bigl(I-\widehat{\widetilde\varphi}(\omega)\bigr)^{-1}\right]^*.
\end{equation}
If, in addition,
\begin{equation}\label{eq:hawkesadmissible}
\widetilde\varphi_{ij}(u)\ge0\ \text{a.e.},
\qquad
\rho(\widetilde\Phi)<1,
\qquad
\widetilde\mu:=(I-\widetilde\Phi)\Lambda\ge0,
\end{equation}
then there exists a stationary ordinary linear Hawkes process with kernel $\widetilde\varphi$ and baseline $\widetilde\mu$ having the same population mean and Bartlett spectrum.
\end{theorem}
The first part is a spectral representation statement, not an existence statement for a signed-intensity point process. Exact pairs satisfying \eqref{eq:hawkesadmissible} are enough to show that population first- and second-order information does not generally identify common information separately from structural excitation on the union of these model classes. Corollary~\ref{cor:oned} gives a univariate example, and Corollary~\ref{cor:exactdelay} gives a bivariate family with directional, nonnegative cross-excitation. The result does not say that every stationary spectrum is generated by a genuine Hawkes process, that every parameter point is observationally equivalent to one, or that second-order identification fails inside every restricted parametric family.

The diagonal-plus-rank-one term in \eqref{eq:spectrum} is not observed separately: it is surrounded by an unknown dynamic transfer function and its adjoint. A static factor decomposition of the observed spectrum therefore cannot by itself select the structural kernel. The equivalence established here concerns means and covariances, not equality of the full point-process laws. This leaves room for higher-order information or restrictions on the transfer function to distinguish the mechanisms. In particular, equal candidate spectra give identical fixed-candidate Whittle contrasts on every record, but empirical periodograms need not have identical sampling distributions. Their variability can depend on higher-order cumulants. The failure asserted here is noninjectivity of the population moment map, not an impossibility result for every procedure computed from event data.

\section{Identification}\label{sec:identification}
There are three distinct ways to obtain more information: use higher-order cumulants, restrict the temporal form of the model, or observe a proxy for the common signal. They support different conclusions. The first gives a local identification theorem for integrated parameters; the second leads to the spectral estimator; the third quantifies residual confounding under an exclusion restriction.

\subsection{Integrated cumulants with Cox immigration}
Let
\[
C=\lim_{T\to\infty}T^{-1}\Cov(N_T,N_T),
\qquad
K_{ijk}=\lim_{T\to\infty}T^{-1}\cum(N_T^i,N_T^j,N_T^k),
\qquad
N_T=N(0,T].
\]
Assume additionally
\begin{equation}\label{eq:cumass}
\E|Y_0|^3<\infty,
\qquad
\int_\R |\Cov(Y_0,Y_u)|\,\dd u<\infty,
\qquad
\int_{\R^2}|\cum(Y_0,Y_u,Y_w)|\,\dd u\dd w<\infty.
\end{equation}
Define
\[
\sigma^2=\int_\R \Cov(Y_0,Y_u)\,\dd u,
\qquad
s_3=\int_{\R^2}\cum(Y_0,Y_u,Y_w)\,\dd u\dd w.
\]
For a two-state chain on $\{0,1\}$ with stationary mean $\bar y$ and total switching rate $\kappa$, $\sigma^2=2\bar y(1-\bar y)/\kappa$. The point of the next proposition is to see what changes once we move beyond second-order information: integrated third-order cumulants retain a different imprint of structural excitation and the common factor.

\begin{proposition}[Integrated cumulants]\label{prop:cumulants}
Under Assumption~\ref{ass:basic} and \eqref{eq:cumass}, let
\[
R=(I-\Phi)^{-1},\qquad u=R\beta,\qquad C^H=R D_\Lambda R^\T,\qquad D_u=R\diag(u)R^\T.
\]
Then
\begin{equation}\label{eq:c2}
C=C^H+\sigma^2uu^\T,
\end{equation}
and
\begin{equation}\label{eq:c3}
K_{ijk}=K^H_{ijk}+\sigma^2\bigl[(D_u)_{ij}u_k+(D_u)_{ik}u_j+(D_u)_{jk}u_i\bigr]+s_3u_iu_ju_k,
\end{equation}
where
\begin{equation}\label{eq:KH}
K^H_{ijk}=\sum_m\Bigl(
R_{im}R_{jm}C^H_{km}+R_{im}C^H_{jm}R_{km}+C^H_{im}R_{jm}R_{km}-2\Lambda_mR_{im}R_{jm}R_{km}
\Bigr).
\end{equation}
These quantities depend on the response measures only through their unit masses and on the excitation kernels only through $\Phi$.
\end{proposition}

\subsection{Local identification from integrated cumulants}
Proposition~\ref{prop:cumulants} gives observable second- and third-order moments as functions of the structural quantities $(\Phi,\beta,\sigma^2,s_3)$. Before asking whether these moments identify the parameters, one scale ambiguity must be removed. The scale of the latent signal is arbitrary: multiplying its loading vector by $c$ and rescaling its long-run cumulants according to
\[
(\beta,\sigma^2,s_3)
\mapsto
(c\beta,\sigma^2/c^2,s_3/c^3)
\]
leaves \eqref{eq:c2}--\eqref{eq:c3} unchanged. We therefore restrict to $\sigma^2>0$ and normalize $\sigma^2=1$.

After this normalization, the finite-dimensional unknowns are the branching matrix $\Phi$, the loading vector $\beta$, and the scalar third cumulant $s_3$. The mean intensity $\Lambda$ is determined by the observed process law, so we treat it as fixed in the identification argument; in estimation it must be learned from data. We write
\[
\theta=(\operatorname{vec}\Phi,\beta,s_3),
\qquad
M_\Lambda(\theta)=(\vech C,\operatorname{vec}_{\mathrm{sym}}K).
\]
Here $M_\Lambda$ is the moment map from structural parameters to the distinct entries of the integrated covariance matrix $C$ and the symmetric third-order cumulant tensor $K$. Local identification asks whether two nearby parameter values can generate the same $(C,K)$. By the inverse-function logic, a sufficient local condition is that the Jacobian $\partial_\theta M_\Lambda(\theta)$ have full column rank.

The dimension count is favorable. For $d=2$, $\theta$ has $4+2+1=7$ coordinates, while $C$ contributes three distinct entries and $K$ contributes four, for seven moments in total. For $d=3$, there are $9+3+1=13$ parameters and $6+10=16$ distinct moments. The next result shows that this counting argument is generically borne out by the Jacobian.

\begin{theorem}[Identification from third-order cumulants]\label{thm:id}
For $d\in\{2,3\}$, the Jacobian $\partial_\theta M_\Lambda(\theta)$ has full column rank $d^2+d+1$ outside a proper algebraic exceptional set in the joint variables $(\Lambda,\theta)$, restricted to the realizable interior with $\Lambda>0$, entrywise $\Phi>0$, $\beta>0$, $\rho(\Phi)<1$, and $\bar\nu=(I-\Phi)\Lambda=\mu+\beta\E Y$ for a nonnegative factor satisfying \eqref{eq:cumass} and a strictly positive baseline $\mu$. Consequently, for almost every positive mean vector $\Lambda$, the normalized finite-dimensional parameter $\theta$ is generically locally identified from $(\Lambda,C,K)$.

For $d=2$ there are seven distinct second- and third-order moments for seven parameters. For $d=3$ there are sixteen distinct moments for thirteen parameters.
\end{theorem}

The conclusion is local rather than global: it rules out observationally equivalent parameter values in a neighborhood of a generic point, but it does not exclude distant solutions. Nor does it identify the shapes $\varphi_{ij}(\cdot)$ or the response measures $g_i$, the delays, the full law of $Y$, or $\mu$ and $\E Y$ separately. At $\beta=0$ the Jacobian is singular because the common-factor contribution begins at second order in $\beta$.

The phrase ``generic'' means that rank failure is confined to a lower-dimensional algebraic exceptional set rather than occurring throughout the parameter space. The proof exploits the fact that the entries of $M_\Lambda(\theta)$ are rational functions of the parameters on the stability region. Hence every maximal Jacobian minor is rational; exhibiting one realizable parameter point at which one such minor is nonzero is enough to show that the same minor is nonzero away from a proper algebraic set. Appendix~\ref{app:certificatesMain} carries out this check in exact rational arithmetic for $d=2$ and $d=3$. The witnesses also belong to an open realizable family: a two-state chain can vary its long-run third cumulant around zero while retaining unit long-run variance and strictly positive baselines. Thus the rank certificate is not obtained only at an inadmissible algebraic point.

\begin{remark}[Weak identification]
Full rank is an identification statement, not a conditioning statement. After row and column normalization, the ratio of the smallest to the largest singular value of the Jacobian has median about $3\times10^{-4}$ across the random $d=2$ designs considered here. The weakest direction loads mainly on $s_3$. Thus the model can be locally identified in population while remaining statistically difficult to estimate from finite samples; Section~\ref{sec:cumest} shows the practical consequence.
\end{remark}

\subsection{Spectral information under parametric restrictions}
The full spectrum \eqref{eq:spectrum} carries more information than its value at $\omega=0$ when the kernels are parametrized. With exponential excitation kernels $\varphi_{ij}(u)=\Phi_{ij}\eta e^{-\eta u}$, a two-state Markov factor, and pure delays, the unknowns are $\theta=(\Phi,\eta,c,\kappa,\Delta)$ with $c_i=\beta_i\sqrt v$ and $\Delta=\delta_2-\delta_1$. This $\theta$ is the spectral parameter vector, distinct from the normalized cumulant parameter in Section~5.2. For $d=2$ it has nine coordinates. The following numerical calculation examines its local rank in representative designs; it does not replace an identification theorem for this family.

\begin{numericalresult}
For $d=2$ and 50 random parameter points, the Jacobian of the binned spectral density in Section~\ref{sec:estimation} with respect to $\theta$, evaluated on 400 frequencies, has full column rank. After column normalization its smallest-to-largest singular value ratio has median $3\times10^{-2}$.
\end{numericalresult}
Information about these parameters is carried by the distinct persistence timescales $1/\kappa$ and $1/\eta$ and by the phase $e^{-i\omega\Delta}$ induced by the response lag. Neither feature alone guarantees injectivity. The singular-value diagnostics for the spectral and cumulant maps use different parameterizations and normalizations, so their numerical ratios are not an invariant comparison of identification strength or statistical efficiency. Theorem~\ref{thm:id} is a separate identification result using $(C,K)$; it does not establish identification of a spectrum-only Whittle estimator. Numerical Result~1 is a local numerical rank diagnostic, not a proof of global parametric identification. The practical spectral estimator is only as reliable as its parametric family. Section~\ref{sec:empirical} shows that improvements in low-frequency fit need not yield a stable decomposition of common variation and integrated excitation.

\subsection{Observed proxies and residual confounding}\label{sec:proxy}
Suppose the econometrician also observes a point process $Z$ whose intensity loads on the same signal,
\begin{equation}\label{eq:proxy}
\lambda_t^Z=\mu_Z+\gamma X_t^Z,
\qquad
X_t^Z=\int_0^\infty \kappa_Ze^{-\kappa_Zu}Y_{t-u}\,\dd u,
\qquad
\Lambda_Z=\E\lambda_t^Z=\mu_Z+\gamma\E Y_0.
\end{equation}
Trades in a related, more liquid asset are the leading example. To interpret such a proxy as information rather than another source of feedback, we need an exclusion condition separating its common-factor dependence from direct excitation.

\begin{assumption}[Proxy exogeneity]\label{ass:proxy}
There is no direct structural excitation between $Z$ and $N$ in either direction, and $Z$ has no self-excitation. Conditionally on $Y$, its compensated martingale is orthogonal to that of $N$. The dependence shared by $N$ and $Z$ is the latent environment in \eqref{eq:model} and \eqref{eq:proxy}.
\end{assumption}
For this calculation we return to the exponential responses of Section~\ref{sec:closed}. Augment the state in \eqref{eq:state} with $X^Z-\E Y$, giving $A_Z,Q_Z$. Let $H_Z$ be the corresponding observation matrix and $D_Z=\diag(\Lambda,\Lambda_Z)$. Let $P_Z$ be the stabilizing filter covariance, $K_Z=P_ZH_Z^\T D_Z^{-1}=[K_{Z,N}\ K_{Z,Z}]$, and $F_Z=A_Z-K_ZH_Z$. Let $\bar H=[H\ 0]$ and let $\bar P$ denote the no-proxy filtering covariance on the augmented state. The next result quantifies what a valid proxy buys us: more information about the latent signal reduces the linear-prediction error attributable to that signal, although not necessarily every fitted coefficient entrywise.

\begin{proposition}[Observed proxy]\label{prop:proxy}
Under Assumptions~\ref{ass:basic} and \ref{ass:proxy}, \eqref{eq:expcov}, and exponential responses:
\begin{enumerate}[label=(\roman*)]
\item The $N$-to-$N$ block of the projected kernel for $(N,Z)$ satisfies
\begin{equation}\label{eq:proxyproj}
I-\widehat{\widetilde\varphi}^{\,Z}_{NN}(\omega)
=\bigl(I-\widehat\Psi_N^Z(\omega)\bigr)\bigl(I-\widehat\varphi(\omega)\bigr),
\qquad
\Psi_N^Z(u)=\bar H e^{F_Zu}K_{Z,N},
\end{equation}
and the projected kernel from $Z$ to $N$ is $\bar H e^{F_Zu}K_{Z,Z}$.
\item $P_Z\preceq\bar P$. Consequently,
\begin{equation}\label{eq:proxyloewner}
\Psi_N^Z(0+)D_\Lambda=\bar H P_Z\bar H^\T
\preceq HPH^\T=\Psi(0+)D_\Lambda.
\end{equation}
\item In the exactly instantaneous-proxy setting $X^Z=Y$, use the reduced $(d+1)$-dimensional state $z$ (so that $\bar H$ is replaced by $H$), hold the $N$-system fixed, and let
\[
r_Z=\frac{2\gamma^2v}{\kappa\Lambda_Z},
\qquad
p(r)=\frac{2v}{1+\sqrt{1+r}}.
\]
Then $\tr P_Z\le(d+1)p(r_Z)$, so $P_Z\to0$ and $\Psi_N^Z(0+)\to0$ as $r_Z\to\infty$. Moreover,
\begin{equation}\label{eq:proxyL2}
\int_0^\infty\left\|\Psi_N^Z(u)D_\Lambda^{1/2}\right\|_F^2\dd u
\le \tr(HP_ZH^\T)=O(r_Z^{-1/2}),
\end{equation}
and, when $\varphi\in L^1$ with the remaining parameters fixed,
\[
\|\widetilde\varphi_{NN}^Z-\varphi\|_{L^2}=O(r_Z^{-1/4}).
\]
\end{enumerate}
\end{proposition}
Loewner order in \eqref{eq:proxyloewner} is not entrywise order. Proposition~\ref{prop:proxy} therefore does not claim that every individual cross-kernel or every integrated coefficient decreases monotonically as a proxy is added. Finite noisy proxies do not automatically identify $\Phi$.

\begin{table}[t]
\centering
\caption{Integrated spurious kernel $\bar\Psi^Z$ after conditioning on a proxy with signal-to-noise ratio $r_Z$. Same design as Figure~\ref{fig:kernels}(b), with $\Phi=0$ and $r=25$ per stream; proxy response rate $\kappa_Z=200\kappa$.}
\label{tab:proxy}
\small
\begin{tabular}{lrrrrr}
\toprule
$r_Z$ & 0 (no proxy) & 12.5 & 125 & 1,250 & 12,500\\
\midrule
Fast$\to$slow $\bar\Psi^Z_{21}$ & 0.594 & 0.454 & 0.145 & 0.018 & 0.002\\
Slow$\to$fast $\bar\Psi^Z_{12}$ & 0.217 & 0.153 & 0.034 & 0.002 & 0.000\\
Self, fast $\bar\Psi^Z_{11}$ & 0.604 & 0.451 & 0.131 & 0.014 & 0.001\\
Self, slow $\bar\Psi^Z_{22}$ & 0.320 & 0.244 & 0.077 & 0.010 & 0.001\\
\bottomrule
\end{tabular}
\end{table}

\begin{remark}[Two proxy failure modes]
First, if $Z$ is itself excited by $N$, Assumption~\ref{ass:proxy} fails and the proxy can absorb structural feedback. Reverse-direction fits by timescale are sensitivity diagnostics, not tests that establish exogeneity: a shared latent signal can itself generate apparent $N\to Z$ excitation. Second, a direct structural response to the proxy's events violates the other exclusion and can mix that response with common information. With a finite noisy proxy, a decline in within-$N$ cross-excitation is not, without additional assumptions, an upper or lower bound on the total information share. Proposition~\ref{prop:proxy} concerns linear-projection errors, not such an attribution bound or a constrained-MLE comparison.
\end{remark}

\section{Estimation and inference}\label{sec:estimation}
We now estimate the restricted spectral specification, rather than invert the integrated cumulants. Its identifying restrictions are maintained explicitly below. Observe $N$ on $[0,T]$, with $T=nh$, and let $X_t=N((th,(t+1)h])$, $t=0,\ldots,n-1$, denote the observed counts in bins of width $h$. Here $X_t$ denotes bin counts, not the latent response states used locally in Section~\ref{sec:closed}. For a Bartlett spectrum $S$, the discrete-time spectral density of $X$ at $\lambda\in(-\pi,\pi]$ is
\begin{equation}\label{eq:binned}
f_h(\lambda;\theta,\Lambda)=hD_\Lambda+\frac1h\sum_{m\in\mathbb Z}
\left[S(\omega_m;\theta)-D_\Lambda\right]
\left(\frac{2\sin(\omega_mh/2)}{\omega_m}\right)^2,
\qquad
\omega_m=\frac{\lambda+2\pi m}{h}.
\end{equation}
The Poisson part aliases exactly to $hD_\Lambda$. For a multivariate exponential kernel, individual off-diagonal alias summands can decay only as $|\omega_m|^{-3}$; the leading odd term cancels upon symmetric pairing of large positive and negative aliases. The structural-recovery simulations use $|m|\le1$, whereas the interior inference experiment evaluates the exact count spectrum. Approximation error must be checked separately from statistical identification. Let
\[
J_k=n^{-1/2}\sum_t(X_t-\bar X)e^{-i\lambda_kt},
\qquad
\lambda_k=2\pi k/n.
\]
We define the Whittle criterion by
\begin{equation}\label{eq:whittle}
\ell_n(\theta,\Lambda)=\sum_{k=1}^{\lfloor n/2\rfloor-1}
\left\{\log\det f_h(\lambda_k;\theta,\Lambda)+\tr\bigl[f_h(\lambda_k;\theta,\Lambda)^{-1}J_kJ_k^*\bigr]\right\}.
\end{equation}
This frequency-domain contrast compares the model spectrum with the observed cross-periodogram; it is not the exact likelihood of the event times. The estimator minimizes $\ell_n(\theta,\widehat\Lambda)$ with $\widehat\Lambda=\bar X/h$, the sample mean count divided by bin width. At nonzero Fourier frequencies, subtracting $\bar X$ does not change $J_k$ because the Fourier weights sum to zero; estimating $\Lambda$ inside the model spectrum is a separate first-order effect.

\subsection{A restricted asymptotic justification}\label{sec:whittleTheory}
The projection and identification results above allow broad latent processes. The asymptotic theory for the Whittle estimator is deliberately narrower. It covers the parametric family used in Section~\ref{sec:simulation}: a bounded Markov information process, bounded response delays, and exponential excitation. This restriction supplies the dependence control that second-order covariance assumptions alone do not provide. We collect these regularity conditions first so that the subsequent probabilistic and Whittle arguments can be stated cleanly.

\begin{assumption}[Regular estimation family]\label{ass:whittle}
Fix $h>0$ and $d$. The stationary data-generating process satisfies Assumption~\ref{ass:basic} and the following conditions.
\begin{enumerate}[label=(\roman*)]
\item $Y$ is an irreducible finite-state continuous-time Markov chain taking bounded nonnegative values, with a strictly positive spectral gap. Responses are pure delays $g_i=\delta_{\delta_i}$, $0\le\delta_i\le D<\infty$.
\item The kernels are $\varphi_{ij}(u)=\Phi_{ij}\eta e^{-\eta u}\1_{\{u>0\}}$. The spectral parameter ranges over a compact set $\Theta$, with $\eta$ bounded above and away from zero and $\rho(\Phi)\le1-\epsilon$. The mean ranges over a compact box $\mathcal L\subset(0,\infty)^d$ whose interior contains $\Lambda_0$. The correctly specified spectrum satisfies $f_h(\lambda;\theta_0,\Lambda_0)=f_0(\lambda)$ for some $\theta_0\in\Theta$.
\item On $\mathcal K=\Theta\times\mathcal L$, $f_h$ is jointly continuous, Hermitian, obeys $f_h(-\lambda)=f_h(\lambda)^\T$, and has eigenvalues uniformly bounded below by a positive constant. Write $B=f_h^{-1}$ and
\[
 B_r(\vartheta)=\frac1{2\pi}\int_{-\pi}^{\pi}B(\lambda;\vartheta)e^{ir\lambda}\dd\lambda,
 \qquad \vartheta=(\theta,\Lambda).
\]
The global envelope satisfies $\sum_r\sup_{\vartheta\in\mathcal K}\|B_r(\vartheta)\|<\infty$. For the normal limit, $f_h$ is jointly $C^2$ in $\vartheta$ on a neighborhood $\mathcal U$ of the truth and, for every multi-index $q$ with $|q|\le2$,
\[
 \sum_{r\in\mathbb Z}(1+|r|)
 \sup_{\vartheta\in\mathcal U}\|\partial_\vartheta^q B_r(\vartheta)\|<\infty.
\]
\end{enumerate}
The estimator uses $\widehat\Lambda=\bar X/h$ when it belongs to $\mathcal L$, and its projection onto $\mathcal L$ otherwise; the modification has probability tending to zero. Factor-law features not identified by the spectrum are fixed in the sampling law and do not enter $\theta$.
\end{assumption}

For the two-state specification used in the simulations one may take $\theta=(\operatorname{vec}\Phi,\eta,c,\kappa,\delta_2,\ldots,\delta_d)$ after fixing one delay origin, with $c_i=\beta_i\sqrt{\operatorname{Var}Y}$. Assumption~\ref{ass:whittle}(iii) is the matrix analogue of the spectral regularity conditions used by \citet{CheyssonLang2022}; Online Supplement Section S.1 verifies the common Fourier envelopes on pole-separated compact bivariate two-state domains, including the regular interior Monte Carlo domain. Corollary~\ref{cor:whittleApprox} records the alias-tail bounds for that two-state parametrization. The first step is to verify that the latent Cox environment does not destroy the short-memory properties required by frequency-domain asymptotics.

\begin{lemma}[Mixing and count moments]\label{lem:coxmix}
Under Assumption~\ref{ass:whittle}(i)--(ii), the stationary bin-count sequence $X_t=N((th,(t+1)h])$ is geometrically strongly mixing: there exist $C,c>0$ such that
\[
\alpha_X(k)\le C e^{-ckh},\qquad k\ge1.
\]
Moreover $\E\|X_0\|^p<\infty$ for every finite $p$. Consequently the reduced count cumulants of every fixed order are absolutely summable with any fixed polynomial lag weight; in particular the second-, third-, and fourth-order cumulant sums needed for the Whittle limit theory are finite.
\end{lemma}

Define the normalized criterion $Q_n(\theta,\Lambda)=n^{-1}\ell_n(\theta,\Lambda)$, where \eqref{eq:whittle} is evaluated with the exact spectrum \eqref{eq:binned}. Its population limit is
\[
Q(\theta,\Lambda_0)=\frac1{4\pi}\int_{-\pi}^{\pi}
\left\{\log\det f_h(\lambda;\theta,\Lambda_0)
+\tr\left[f_h(\lambda;\theta,\Lambda_0)^{-1}f_0(\lambda)\right]\right\}\dd\lambda,
\]
where $f_0(\lambda)=f_h(\lambda;\theta_0,\Lambda_0)$. Let
\[
\Theta_0=\{\theta\in\Theta:f_h(\lambda;\theta,\Lambda_0)=f_0(\lambda)\text{ for a.e. }\lambda\}.
\]
This set is kept explicit because Theorem~\ref{thm:id} concerns $(\Lambda,C,K)$, not spectrum-only identification. A unique spectral target, or selection of a consistent local fit near one isolated target, is an additional requirement; a small optimization residual does not establish either.

At an interior identified point, write $f_a=\partial_{\theta_a}f_h(\cdot;\theta_0,\Lambda_0)$ and $f_{\Lambda_i}=\partial_{\Lambda_i}f_h(\cdot;\theta_0,\Lambda_0)$, and define
\begin{align}
H_{ab}&=\frac1{4\pi}\int_{-\pi}^{\pi}\tr(f_0^{-1}f_a f_0^{-1}f_b)\dd\lambda,\label{eq:whittleH}\\
A_{ai}&=\frac1{4\pi}\int_{-\pi}^{\pi}\tr(f_0^{-1}f_a f_0^{-1}f_{\Lambda_i})\dd\lambda.\label{eq:whittleA}
\end{align}
In this section $H$ is the curvature of the expected contrast and $A$ measures how its parameter score changes with the mean intensity. These roles differ from those of the state-space matrices in Section~\ref{sec:closed}. Let $G_n=\sqrt n\,\partial_\theta Q_n(\theta_0,\Lambda_0)$ and $U_n=\sqrt n(\widehat\Lambda-\Lambda_0)$. Lemma~\ref{lem:coxmix} and Assumption~\ref{ass:whittle}(iii) imply the joint central limit theorem
\begin{equation}\label{eq:GUclt}
\binom{G_n}{U_n}\Rightarrow N\!\left(0,
\begin{bmatrix}\Omega_{GG}&\Omega_{G\Lambda}\\
\Omega_{\Lambda G}&\Omega_{\Lambda\Lambda}\end{bmatrix}\right),
\qquad \Omega_{\Lambda\Lambda}=h^{-2}f_0(0).
\end{equation}
The covariance is defined by the absolutely convergent long-run covariance of the time-domain score and mean in Online Supplement Section S.1. Its block $\Omega_{GG}$ contains a fourth-order cumulant correction, while $\Omega_{G\Lambda}$ contains third-order cumulants. With the dependence control and score limit in place, we can now state the consistency and interior asymptotic-normality result for the Whittle estimator.

\begin{theorem}[Whittle estimation with Cox immigration]\label{thm:whittle}
Under Assumption~\ref{ass:whittle} and Lemma~\ref{lem:coxmix}:
\begin{enumerate}[label=(\roman*)]
\item If $\widehat\theta_n$ is an approximate global minimizer of $Q_n(\theta,\widehat\Lambda)$ with objective error $o_p(1)$, then
\[
\operatorname{dist}(\widehat\theta_n,\Theta_0)\xrightarrow{p}0.
\]
In particular, $\widehat\theta_n\xrightarrow{p}\theta_0$ if $\Theta_0=\{\theta_0\}$.
\item Suppose $\theta_0$ is an interior isolated member of $\Theta_0$, $H$ in \eqref{eq:whittleH} is nonsingular, and $\widehat\theta_n$ is a consistent local minimizer whose score residual is $o_p(n^{-1/2})$. Then
\begin{equation}\label{eq:whittleCLT}
\sqrt n(\widehat\theta_n-\theta_0)
\Rightarrow N(0,V_{\mathrm{eff}}),
\end{equation}
where
\begin{equation}\label{eq:whittleV}
V_{\mathrm{eff}}=H^{-1}
\left(\Omega_{GG}+A\Omega_{\Lambda\Lambda}A^\T
+\Omega_{G\Lambda}A^\T+A\Omega_{\Lambda G}\right)H^{-\T}.
\end{equation}
\end{enumerate}
Thus estimating the mean intensity inside the model spectrum is generally a first-order nuisance correction; the variance in \eqref{eq:whittleV} need not equal the Gaussian inverse-information matrix.
\end{theorem}

The theorem concerns an interior regular fit. It does not cover the null $c=0$, where factor persistence and delay are unidentified and the loading enters the spectrum quadratically, nor does it give normal limits for zero cross-excitation coefficients on the boundary. The no-imitation design in Section~\ref{sec:simulation} is therefore a mechanism experiment rather than a coverage experiment. The exact aliased spectrum is inconvenient computationally, so the next corollary records when truncating the alias sum and pooling nearby frequencies leave these large-sample conclusions unchanged.

\begin{corollary}[Alias truncation and frequency pooling]\label{cor:whittleApprox}
Let $f_h^{(M)}$ denote \eqref{eq:binned} with aliases restricted symmetrically to $|m|\le M$. For the two-state spectral parametrization on a compact set satisfying Assumption~\ref{ass:whittle}, with $c$ bounded and $\kappa$ bounded above and away from zero,
\[
\sup_{\lambda,\theta}\left(\|f_h^{(M)}-f_h\|+\|\partial f_h^{(M)}-\partial f_h\|\right)=O(M^{-2}),
\qquad
\sup_{\lambda,\theta}\|\partial^2 f_h^{(M)}-\partial^2 f_h\|=O(M^{-1}),
\]
where derivatives include the mean-intensity nuisance. If $M_n\to\infty$, alias truncation preserves the consistency conclusion of Theorem~\ref{thm:whittle}; if in addition $\sqrt n\,M_n^{-2}\to0$, it preserves the first-order limit \eqref{eq:whittleCLT}. If adjacent Fourier frequencies are pooled using block-size weights and maximal frequency width $w_n$, the corresponding sufficient conditions are $w_n\to0$ for consistency and $\sqrt n\,w_n\to0$ for the same first-order limit. In particular, pooling a fixed number of adjacent Fourier frequencies satisfies these conditions.
\end{corollary}

The structural-recovery simulations use the computationally cheaper fixed truncation $|m|\le1$; this is a finite-sample numerical approximation rather than the asymptotic sequence in Corollary~\ref{cor:whittleApprox}. In the true-imitation benchmark, increasing the truncation produces changes in the population spectrum that are negligible relative to sampling error; details are reported in the Online Supplement. Pure-Hawkes count Whittle theory is developed by \citet{CheyssonLang2022}; recent multivariate event-time theory is given by \citet{TangEtAl2026}. Lemma~\ref{lem:coxmix} is the extra dependence step required by the random Cox environment.

\subsection{Feasible sandwich standard errors}\label{sec:feasibleSE}
The covariance in Theorem~\ref{thm:whittle} can be estimated from a single count record without estimating a trispectrum or specifying the latent factor's higher-order law. We use the time-domain effective score and a positive-semidefinite Bartlett heteroskedasticity-and-autocorrelation-consistent (HAC) estimator \citep{NeweyWest1987,Andrews1991}. This is a feasible implementation of the sandwich limit, not an assumption of independent periodogram ordinates.

Let $\widehat f=f_h(\cdot;\widehat\theta,\widehat\Lambda)$ and evaluate $\widehat H,\widehat A$ by substituting the fitted values in \eqref{eq:whittleH}--\eqref{eq:whittleA}. Define
\[
 \widehat W_a(\lambda)=\widehat f^{-1}\widehat f_a\widehat f^{-1},\qquad
 \widehat f_a=\left.\partial_{\theta_a}f_h\right|_{(\widehat\theta,\widehat\Lambda)},
 \qquad
 \widehat W_{a,r}=\frac1{2\pi}\int_{-\pi}^{\pi}\widehat W_a(\lambda)e^{ir\lambda}\dd\lambda.
\]
Write $Z_t=X_t-\bar X$, take a filter radius $R_n$, and use $\mathcal T_n=\{R_n,\ldots,n-1-R_n\}$, with $N_n=n-2R_n$. For $t\in\mathcal T_n$ set
\begin{equation}\label{eq:feasibleScore}
 v_{a,t}=-\frac12\sum_{|r|\le R_n}Z_t^\T\widehat W_{a,r}Z_{t-r},
 \qquad
 \widehat q_t=v_t-\bar v+\widehat A Z_t/h,
\end{equation}
where $\bar v$ averages over $\mathcal T_n$. Center $\widehat q_t$ over the same index set and denote the result by $\widetilde q_t$. The minus-lag orientation in \eqref{eq:feasibleScore} matches the Fourier convention in Online Supplement Section S.1. A deterministic model-centering term is unnecessary because the HAC input is sample-centered.

Reindex these $N_n$ vectors from zero. For an integer bandwidth $b_n$ define
\[
 \widehat\Gamma_q(k)=\frac1{N_n}\sum_{t=k}^{N_n-1}\widetilde q_t\widetilde q_{t-k}^{\T},
\]
\begin{equation}\label{eq:feasibleMeat}
 \widehat\Omega_q=\widehat\Gamma_q(0)+
 \sum_{k=1}^{b_n-1}\left(1-\frac{k}{b_n}\right)
       \{\widehat\Gamma_q(k)+\widehat\Gamma_q(k)^\T\},
 \qquad
 \widehat V=\widehat H^{-1}\widehat\Omega_q\widehat H^{-\T}.
\end{equation}
The covariance estimator for $\widehat\theta$ is $\widehat V/n$, not $\widehat V/N_n$: the point estimate uses the original $n$ observations. The HAC meat automatically includes the score--mean cross-covariances in \eqref{eq:whittleV}. Omitting $\widehat A Z_t/h$ instead estimates a different, generally incorrect variance for the two-step estimator. The remaining question is whether this single-record HAC construction consistently estimates the covariance appearing in Theorem~\ref{thm:whittle}; the next proposition gives sufficient conditions.

\begin{proposition}[Feasible interior inference]\label{prop:feasibleSE}
Assume Theorem~\ref{thm:whittle}(ii) and, locally at the truth, common exponential Fourier envelopes for $W_a$ and its first parameter--mean derivatives. These envelopes hold on the compact pole-separated family in Online Supplement Section S.1. Let
\[
 b_n\longrightarrow\infty,\quad b_n=o(\sqrt n),\quad R_n\longrightarrow\infty,
 \quad R_n=o(n),\quad b_ne^{-aR_n}\longrightarrow0
\]
for a common envelope exponent $a>0$. If numerical errors have a deterministic local uniform envelope with $\sum_{|r|\le R_n}\sup_{\vartheta\in\mathcal U}\|W^{\rm num}_{a,r}(\vartheta)-W_{a,r}(\vartheta)\|+\sup_{\mathcal U}\|A^{\rm num}-A\|\le\epsilon_n$, also require $b_n\epsilon_n\to0$; the numerical $H$ error must vanish. Then
\[
 \widehat V\xrightarrow{p}V_{\rm eff}.
\]
For every coordinate with $(V_{\rm eff})_{aa}>0$,
\[
 \frac{\sqrt n(\widehat\theta_a-\theta_{0a})}{\sqrt{\widehat V_{aa}}}
 \ \Rightarrow\ N(0,1).
\]
Thus $\widehat\theta_a\pm z_{1-\alpha/2}\sqrt{\widehat V_{aa}/n}$ has asymptotic marginal coverage $1-\alpha$. Smooth reparametrizations, including log coordinates for positive parameters, follow by the delta method. These are pointwise interior conclusions, not uniform guarantees near boundaries or weakly identified parameter sequences.
\end{proposition}

For numerical conditioning the implementation works in log coordinates and transforms the fitted covariance back to physical coordinates by the delta method. The Monte Carlo uses deterministic filter and HAC bandwidth sequences satisfying Proposition~\ref{prop:feasibleSE}; numerical choices and robustness checks are collected in Online Supplement Section S.2. No eigenvalue clipping is used.

\subsection{Scope of the asymptotic result}\label{sec:implementScope}
Theorem~\ref{thm:whittle} and Proposition~\ref{prop:feasibleSE} concern a single increasing stationary record at an interior regular point. They do not cover the no-factor null, zero cross-excitation coefficients on the boundary, or local-to-critical sequences. The empirical comparison in Section~\ref{sec:empirical} instead pools fixed-hour periodograms and fits their expected finite-window values, accounting for the corresponding Fej\'er smoothing. Matching that target does not extend the theorem to fixed-window pooled estimates or to solutions on loading and immigration boundaries. We therefore report the SOL fits descriptively, without applying these interior confidence intervals. Online Supplement Sections S.3.1 and S.3.3 give the finite-window formula and empirical implementation.

\section{Simulation evidence}\label{sec:simulation}
The first experiment compares two exactly second-order-equivalent mechanisms and isolates the information supplied by a higher-order contrast. The subsequent experiments address different questions: misspecified Hawkes fitting, recovery under a restricted spectral family, and calibration of regular interior inference. Their distinct assumptions and targets are kept separate.

\subsection{An exactly second-order-equivalent pair}\label{sec:equivalentPairMC}
We first isolate the information gained by moving beyond the spectrum. In the pure-delay construction of Corollary~\ref{cor:exactdelay}, let $Y$ be a stationary symmetric $\{0,1\}$ chain with total switching rate one, and set $\lambda_t^{C,1}=1+2Y_t$ and $\lambda_t^{C,2}=1+2Y_{t-1/2}$. This process has no structural excitation. Its genuine Hawkes counterpart has exactly the same mean and Bartlett spectrum, but an integrated excitation matrix with spectral radius approximately $0.4022$. Thus any Whittle criterion evaluated at these two fixed candidate spectra is identical on every record. This statement concerns the candidate spectra, not equality of the sampling distributions of periodogram estimates.

The two mechanisms differ beyond second order. The following calculation isolates that difference without replacing a finite-window moment by a long-run limit.

\begin{proposition}[Finite-window separation of the symmetric pair]\label{prop:finitewindow}
Let $Y$ be a stationary symmetric $\{0,1\}$ Markov chain with total switching rate $\kappa>0$. Consider the Cox intensities $\mu+\beta Y_t$ and $\mu+\beta Y_{t-\Delta}$, with $\mu\ge0$, $\beta>0$ and $\Delta>0$, and their Hawkes counterpart from Corollary~\ref{cor:exactdelay}. For either process put $Q_H=N^1((0,H])+N^2((0,H])$ and define its third factorial cumulant by
\[
\mathcal F_3(H)=\operatorname{cum}_3(Q_H)-3\operatorname{Var}(Q_H)+2\E Q_H.
\]
For every $H>0$, the Cox process has $\mathcal F_3^C(H)=0$, whereas the Hawkes counterpart has $\mathcal F_3^H(H)>0$.
\end{proposition}
\begin{proof}
Conditional on $Y$, the Cox total is Poisson with parameter
\[
A_H=2\mu H+\beta\left(\int_0^H Y_t\,\dd t+\int_0^H Y_{t-\Delta}\,\dd t\right).
\]
Its third factorial cumulant equals $\operatorname{cum}_3(A_H)$. The paths $Y$ and $1-Y$ have the same law, so $A_H-\E A_H$ is symmetric and that cumulant vanishes.

For the Hawkes process, let $Z_{j,s}(H)$ be the number of points in $(0,H]$ from the cluster rooted at a type-$j$ immigrant at $s$. The Poisson-cluster representation \citep{HawkesOakes1974} gives
\[
\mathcal F_3^H(H)=\sum_j\mu_j^H\int_{\R}\E\bigl[Z_{j,s}(H)(Z_{j,s}(H)-1)(Z_{j,s}(H)-2)\bigr]\,\dd s.
\]
This quantity is finite by subcriticality and finite third moments of Poisson-offspring cluster sizes: for a cluster of size $S$, $\int Z_s(H)^3\,\dd s\le H S^3$. It is strictly positive because the counterpart baselines are positive and the kernels are positive near zero, giving a positive probability of three cluster points in the window for a positive-measure set of root times.
\end{proof}

The calculation uses the standard mixed-Poisson and cluster identities; its role here is to separate the explicit second-order-equivalent pair at every finite window. Symmetry is essential to the zero null, and the proposition is not a test characterizing all Cox processes.

We estimate the contrast from consecutive length-eight windows and studentize it using a data-based Bartlett HAC variance. Table~\ref{tab:equivalentPairMC} reports rejection at a one-sided 5\% normal cutoff for 512 independent evaluation records per mechanism. Record lengths are nested prefixes of the same independent longest records. A prespecified weaker-signal pair replaces the Cox intensities by $1.5+Y_t$ and $1.5+Y_{t-1/2}$; its mean remains two per stream, and its counterpart radius is approximately $0.1756$.

\begin{table}[htbp]
\centering\small
\caption{Third-order separation of exactly second-order-equivalent pairs. Entries are rejection percentages for the finite-window HAC diagnostic, with 512 independent evaluation records per mechanism and length. Cox entries measure false positives; Hawkes entries measure power. The primary window is $H=8$.}
\label{tab:equivalentPairMC}
\begin{tabular}{llrrrr}\toprule
Signal & Mechanism & $T=2^{10}$ & $2^{12}$ & $2^{14}$ & $2^{16}$\\\midrule
$r=1$ & Symmetric Cox & 3.9 & 4.9 & 4.7 & 4.5\\
& Hawkes counterpart & 23.4 & 68.9 & 100.0 & 100.0\\\addlinespace
$r=0.25$ & Symmetric Cox & 2.9 & 5.1 & 4.1 & 5.3\\
& Hawkes counterpart & 5.7 & 8.2 & 18.4 & 46.1\\\bottomrule
\end{tabular}
\end{table}

The primary pair becomes readily distinguishable at the larger tested lengths, while the weaker pair remains difficult even in the longest records. This connects exact second-order ambiguity to usable but potentially noisy higher-order information. It is not a generic test for the absence of excitation: an asymmetric Cox factor can have a nonzero third factorial cumulant. Nor does the experiment establish recovery of all normalized parameters or apply the interior identification theorem at the boundary. Full results for the prespecified windows and an independent known-design Monte Carlo reference are provided in Online Supplement Section S.2.1.

\subsection{Misspecified Hawkes fitting}\label{sec:misspecified}
In a separate experiment we set $\varphi\equiv0$ and simulate two streams driven only by a symmetric Markov signal $U_t\in\{-1,1\}$,
\[
\lambda_t^1=\Lambda(1+\mathrm{CV}\,U_t),
\qquad
\lambda_t^2=\Lambda(1+\mathrm{CV}\,U_{t-0.3}),
\]
with $\Lambda=50$, covariance decay rate $\kappa=1$, and $T=4000$. This is within Assumption~\ref{ass:basic}: write $Y_t=(U_t+1)/2$, $\mu_i=\Lambda(1-\mathrm{CV})$, and $\beta_i=2\Lambda\mathrm{CV}$. A bivariate Hawkes model with nonnegative sum-of-exponentials kernels is then estimated by maximum likelihood. Table~\ref{tab:mle} compares the fitted branching coefficients with the second-order projection. In this design the MLE approximates both the magnitude and the fast-to-slow direction of the projected kernel: at CV $=0.5$, for example, the projected fast-to-slow coefficient is $0.651$ and the fitted value is $0.639$. The comparison links the linear-prediction calculation to a conventional constrained fit in this example, without asserting that their misspecification targets coincide generally. With eight replications, it is a mechanism illustration rather than a precise assessment of sampling distributions.

\begin{table}[H]
\centering
\caption{Spurious branching under pure common information ($\Phi=0$). Projection entries are from Corollaries~\ref{cor:oned} and \ref{cor:exactdelay}; MLE entries are mean $\pm$ s.d. over eight replications.}
\label{tab:mle}
\small
\begin{tabular}{lcrrrr}
\toprule
&&\multicolumn{2}{c}{CV $=0.25$}&\multicolumn{2}{c}{CV $=0.5$}\\
\cmidrule(lr){3-4}\cmidrule(lr){5-6}
Quantity & Truth & Projection & MLE & Projection & MLE\\
\midrule
Univariate $n$ & 0 & 0.629 & $0.639\pm0.047$ & 0.804 & $0.836\pm0.004$\\
Fast$\to$slow $\widetilde\Phi_{21}$ & 0 & 0.527 & $0.495\pm0.118$ & 0.651 & $0.639\pm0.084$\\
Slow$\to$fast $\widetilde\Phi_{12}$ & 0 & 0.171 & $0.155\pm0.016$ & 0.103 & $0.093\pm0.033$\\
$\rho(\widetilde\Phi)$ & 0 & 0.707 & 0.71 & 0.832 & 0.83\\
\bottomrule
\end{tabular}
\end{table}

\subsection{Recovering structural excitation}\label{sec:recovery}
We next simulate the full model with $d=2$, exponential excitation decay $\eta=2$, $\mu=(0.5,0.5)$, $\beta=(1.5,1.5)$, a two-state Markov factor with $\bar y=0.3$ and $\kappa=0.2$, and response delays $\delta=(0,1)$. We compare
\[
\Phi_{\mathrm{no\ imitation}}=\begin{pmatrix}0.3&0\\0&0.3\end{pmatrix},
\qquad
\Phi_{\mathrm{imitation}}=\begin{pmatrix}0.3&0.05\\0.25&0.3\end{pmatrix}.
\]
Each replication contains about $6$--$7\times10^5$ events. Table~\ref{tab:mc} compares the factor-aware Whittle estimator with pure-Hawkes Whittle and maximum likelihood. The factor-aware estimator recovers the branching matrix, lag and signal persistence in both designs. In contrast, the pure-Hawkes fits exhibit the confounding mechanism described in Section~\ref{sec:spurious}: with no imitation they report fast-to-slow branching of $0.21$--$0.29$; with true $\Phi_{21}=0.25$ they estimate $0.436$ and $0.495$, overstatements of about $75\%$ and $100\%$.

\begin{table}[H]
\centering
\caption{Structural-recovery Monte Carlo. Mean (standard deviation) over 12 replications per design, $T=2\times10^5$, about 676 thousand events.}
\label{tab:mc}
\scriptsize
\begin{tabular}{llrrrr}
\toprule
Design & Parameter & Truth & Factor-aware Whittle & Pure-Hawkes Whittle & Pure-Hawkes MLE\\
\midrule
\multirow{6}{*}{No imitation}
& $\Phi_{11}$ & 0.30 & 0.303 (0.006) & 0.561 (0.002) & 0.579 (0.004)\\
& $\Phi_{12}$ (slow$\to$fast) & 0.00 & 0.003 (0.003) & 0.119 (0.003) & 0.139 (0.003)\\
& $\Phi_{21}$ (fast$\to$slow) & 0.00 & 0.002 (0.002) & 0.210 (0.002) & 0.289 (0.002)\\
& $\Phi_{22}$ & 0.30 & 0.300 (0.004) & 0.529 (0.002) & 0.494 (0.004)\\
& $\Delta$ & 1.00 & 1.021 (0.031) & -- & --\\
& $\kappa$ & 0.20 & 0.200 (0.002) & -- & --\\
\midrule
\multirow{6}{*}{True imitation}
& $\Phi_{11}$ & 0.30 & 0.298 (0.008) & 0.511 (0.002) & 0.533 (0.003)\\
& $\Phi_{12}$ (slow$\to$fast) & 0.05 & 0.048 (0.005) & 0.140 (0.002) & 0.153 (0.002)\\
& $\Phi_{21}$ (fast$\to$slow) & 0.25 & 0.248 (0.009) & 0.436 (0.004) & 0.495 (0.006)\\
& $\Phi_{22}$ & 0.30 & 0.300 (0.006) & 0.460 (0.003) & 0.434 (0.005)\\
& $\Delta$ & 1.00 & 0.979 (0.051) & -- & --\\
& $\kappa$ & 0.20 & 0.202 (0.004) & -- & --\\
\bottomrule
\end{tabular}
\end{table}

On a single longer path with $6.8\times10^6$ events, the factor-aware estimate remains close to the truth, while the pure-Hawkes fit again absorbs common variation into excitation. These results show why a good pure-Hawkes fit is not sufficient evidence of structural interaction and motivate explicit modeling of the common-information channel.

\subsection{Regular interior inference}\label{sec:rootn}\label{sec:coverage}
The no-excitation and no-cross-excitation designs include boundary parameters. To assess sampling rates and interval calibration separately from the misspecification comparisons, we use the following strictly interior design:
\[
 \Phi=\begin{pmatrix}.25&.12\\.22&.30\end{pmatrix},\quad
 \eta=3,\quad c=(1,.8)^\T,\quad \kappa=.2,\quad \Delta=1.15.
\]
The factor is a two-state chain and all nine estimated spectral parameters are strictly interior. We use $h=.25$ and $n\in\{2^{15},2^{17},2^{19}\}$, with 160 independent paths at each length. All nine parameters and the mean intensity are estimated; the fit uses the exact count spectrum. Full simulation and numerical details, including the rate table and optimization diagnostics, are in Online Supplement Section S.2.

\begin{figure}[t]
\centering
\includegraphics[width=.99\textwidth]{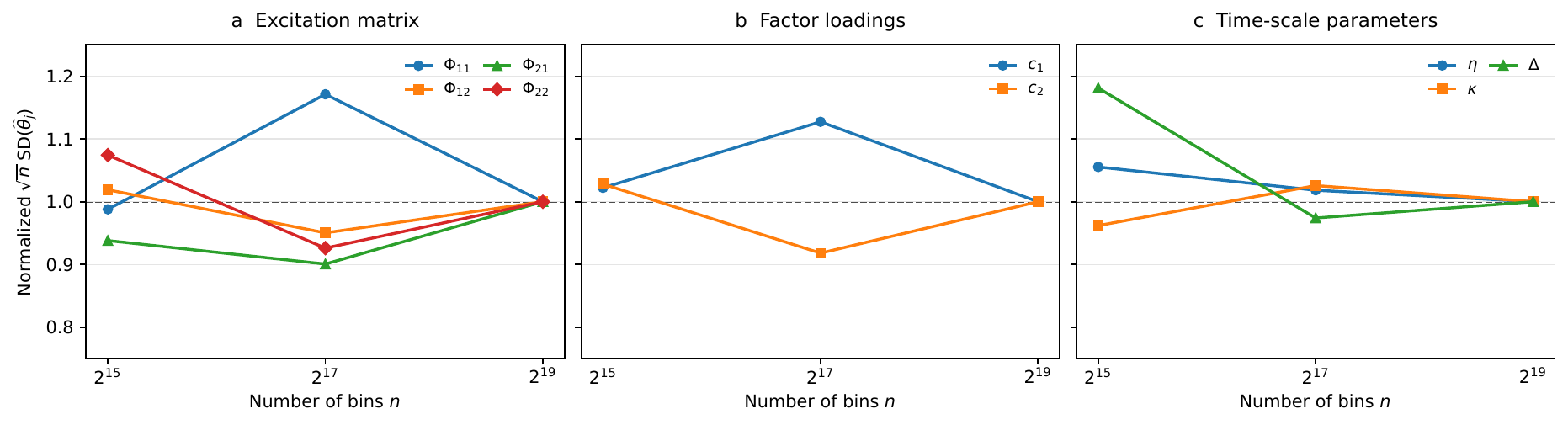}
\caption{Monte Carlo check of the root-$n$ dispersion rate for the nine interior Whittle parameters (160 independent records at each $n$). Each curve is $\sqrt{n}\,\mathrm{SD}(\widehat\theta_j)$ divided by the same quantity at $n=2^{19}$, computed from the simulated estimates. Under exact root-$n$ scaling the curve is horizontal at one (dashed reference line). All curves equal one at the largest $n$ by construction. Panels separate (a) excitation coefficients, (b) factor loadings, and (c) temporal parameters. This is a rate diagnostic, not a confidence-interval coverage plot.}
\label{fig:rootn}
\end{figure}

Across the three sample sizes, the slopes of $\log\mathrm{SD}(\widehat\theta)$ against $\log n$ range from $-0.560$ to $-0.477$, and Monte Carlo bootstrap intervals for all nine slopes contain the root-$n$ benchmark $-0.5$. To make the rate comparison directly interpretable, Figure~\ref{fig:rootn} plots $\sqrt{n}\,\mathrm{SD}(\widehat\theta_j)$ divided by its value at $n=2^{19}$ for each parameter $j$. If the standard deviation decreases exactly at the $n^{-1/2}$ rate, this ratio is constant at one. An upward or downward departure at shorter records indicates a departure from that rate over the sampled range; the rightmost value is one \emph{by normalization}, not because the experiment verifies the asymptotic theory there. The parameters are grouped into excitation coefficients, factor loadings, and temporal parameters to make each trend visible. The plotted dispersions are broadly compatible with root-$n$ scaling in this design, but three sample sizes do not establish a limiting distribution or global identification. We examine interval calibration separately below.

We then apply Proposition~\ref{prop:feasibleSE} to the frozen 480 fits, using only each record and its fitted parameters. After inspecting the initial coverage results, including short-record delay undercoverage, we add 320 independent confirmation records at $n=2^{19}$ without changing the sampling law, optimizer, bandwidth rules or interval formula. Table~\ref{tab:coverage} reports the original and confirmation results separately.

\begin{table}[H]
\centering
\caption{Marginal 95\% physical-coordinate Wald coverage (percent). The confirmation sample contains 320 new independent records at $n=2^{19}$. Mean SE/SD uses empirical SD only to evaluate the standard errors, never to construct an interval.}
\label{tab:coverage}
\small
\begin{tabular}{lrrrrr}
\toprule
&\multicolumn{3}{c}{Original records, $n$}&\multicolumn{2}{c}{Independent confirmation}\\
\cmidrule(lr){2-4}\cmidrule(lr){5-6}
Parameter & $2^{15}$ & $2^{17}$ & $2^{19}$ & $2^{19}$ & Mean SE/SD\\
\midrule
$\Phi_{11}$ & 93.8 & 89.4 & 96.2 & 92.2 & 0.957 \\
$\Phi_{12}$ & 93.1 & 93.8 & 94.4 & 94.7 & 0.971 \\
$\Phi_{21}$ & 93.1 & 95.0 & 90.6 & 94.7 & 0.997 \\
$\Phi_{22}$ & 93.1 & 96.9 & 94.4 & 92.8 & 0.958 \\
$\eta$ & 94.4 & 93.1 & 95.6 & 92.5 & 0.912 \\
$c_1$ & 93.1 & 93.1 & 96.2 & 93.4 & 0.921 \\
$c_2$ & 90.6 & 94.4 & 92.5 & 94.1 & 0.967 \\
$\kappa$ & 94.4 & 93.1 & 95.6 & 93.8 & 0.966 \\
$\Delta$ & 84.4 & 93.8 & 93.8 & 94.4 & 0.968 \\
\bottomrule
\end{tabular}
\end{table}

The feasible standard errors track dispersion reasonably well: in the independent confirmation, mean SE/SD ratios are $0.912$--$0.997$. The corresponding coverages are $92.2\%$--$94.7\%$, indicating modest anti-conservatism rather than exact nominal calibration. The main finite-sample failure is the delay in the shortest records, where coverage is $84.4\%$; a log-Wald interval raises it only to $88.8\%$. We therefore interpret Proposition~\ref{prop:feasibleSE} as a pointwise interior asymptotic result, not a uniform finite-sample guarantee. The mean-intensity correction is empirically material: for $\eta$ in the original largest-$n$ sample, Gaussian inverse-information standard errors yield $88.8\%$ coverage, while omitting the plug-in correction produces substantial overcoverage; the intended sandwich gives $95.6\%$. Online Supplement Section S.2 reports the corresponding sensitivity and numerical checks.

\subsection{Why third-order identification is not our default estimator}\label{sec:cumest}
Theorem~\ref{thm:id} is an identification result, but its moment map is weakly conditioned. At population values of $(\Lambda,C,K)$, nonlinear least squares recovers the parameter essentially exactly. Replacing those moments by non-overlapping-window estimates on the $6.8\times10^6$-event path produces unstable estimates of $\Phi$, and $\widehat s_3$ often hits the search bounds; third-order window cumulants have sampling error of roughly $10\%$ in this experiment. In this paper the cumulant equations establish identification; practical estimation uses the restricted parametric spectrum. They may also motivate model checks, but we do not report a calibrated over-identification test.

\section{Empirical illustration: fit and structural attribution}\label{sec:empirical}
The exact construction above concerns equality of population covariance measures. A fitted spectrum over a selected frequency range is a different object. We use SOL spot and perpetual-futures activity to illustrate this distinction, not to identify how much of the observed dependence is caused by a latent factor. Detailed implementation and all associated sensitivity analyses are in Online Supplement Section S.3.

\subsection{Recorded activity and matched spectral fits}\label{sec:empiricalData}
The processed Binance records cover August 2026 and contain 4,977,119 SOLUSDT spot events and 7,807,910 USD-M perpetual events. Spot timestamps vary within milliseconds, while perpetual timestamps lie on millisecond ticks. Counts describe aggregate activity, not individual traders or signed order flow. Across 27 distinct timestamp and origin settings with 0.5--2 ms bins, the count cross-correlation peaks at a positive perpetual-to-spot lag of 1--2 ms. Coarser bins and phase-band choices affect the measured magnitude. This supports precedence in recorded activity, not a causal or precisely measured reaction time; the processed files do not establish exchange-clock synchronization.

For the spectral comparison we retain recorded timestamps, all 744 UTC hours, 2 ms bins, hourly demeaning, and 232 logarithmic frequency groups from approximately $1/1200$ to 244 Hz. Each group receives its number of raw Fourier ordinates as weight. Both models have the same nonnegative exponential excitation basis. The factor specification adds four exponential covariance modes with a shared loading vector; zero factor strengths nest the pure model. We impose a spectral-radius cap and the necessary mean-immigration constraint $(I-\Phi)\Lambda\ge0$. The model target is the expected one-hour periodogram, including bin integration and finite-window smoothing \citep{SykulskiEtAl2019}.

With the common six-scale basis, the fitted radii are about $0.9633$ and $0.9588$; raising the cap from $0.99$ to $0.999$ does not change them materially. The factor nevertheless improves low-frequency fit: at a 20-minute period, the spot periodogram relative to the Poisson level is about 1,007, compared with fitted values of 256 and 1,009. Frequencies below 1 Hz carry only about $0.42\%$ of the aggregate criterion weight. Thus a small difference in the overall criterion can hide substantial frequency-local differences. Table~\ref{tab:solwhittle} documents the separate sensitivity of the integrated radius to the common basis.

\begin{table}[htbp]
\centering\small
\caption{SOL illustration: fitted spectral radii under matched excitation bases. Both models use the same basis in each row, cap $0.999$, and nonnegative mean immigration. The additional 0.1 ms mode is an exploratory specification check, not a claim about execution-time precision.}
\label{tab:solwhittle}
\begin{tabular}{lrr}\toprule
Common excitation basis & Pure Hawkes & Factor model\\\midrule
1 ms--1 s (four scales) & 0.9322 & 0.9185 \\
1 ms--100 s (six scales) & 0.9633 & 0.9588 \\
Six scales plus 1000 s & 0.9633 & 0.9990 \\
Six scales plus 0.1 ms & 0.9629 & 0.9303 \\
\bottomrule\end{tabular}
\end{table}

\subsection{Profiles and information outside the fitted frequencies}\label{sec:empiricalProfiles}
We next prescribe $\rho(\Phi)$ within each fixed basis and reoptimize the remaining coordinates. The resulting numerical profile is not flat over the full range $0.70$--$0.999$: substantial reductions in branching have a larger fitting cost. Its upper region is much shallower with the 1000 s mode. Within that seven-scale factor family, radii $0.9588$ and $0.999$ have nearby within-hour criteria, but their lowest-frequency fits are not identical. No likelihood-ratio cutoff or confidence set is inferred from these numerical profiles.

The hourly totals provide second-order information not fitted by that objective. Their observed variance-to-mean ratios are about 10,007 and 6,467. The seven-scale factor fits at radii $0.9588$ and $0.999$ imply very different finite-month sample dispersions: about 1,593 versus 346,286 for spot, and 980 versus 520,434 for perpetual. A candidate near $0.98$ is closer in variance but substantially underpredicts one-hour persistence. Applying the same hour-of-day and weekday mean-removal operation to the data and model covariances does not remove that mismatch. These are diagnostics of the frozen fits, not a rejection of every member of the family; temporal rate variation also prevents a simple stationary confidence interpretation.

The main factor fit has a zero direct perpetual loading and spot mean immigration close to its numerical lower margin. We therefore do not apply the interior standard errors. A separate, retained event-level BTC conditioning comparison is documented in Supplement Section S.3.6; it is neither a refit of these spectral models nor evidence establishing the proxy exclusion. The illustration's conclusion is about the information used by calibration: good within-hour fit alone does not validate a long-run excitation decomposition. The exact equivalence result and the symmetric-pair third-order separation stand independently of this empirical lack of fit.

\section{Conclusion}
Heterogeneous responses to latent common information can create directional Hawkes-form prediction kernels without event-to-event excitation. Building on the classical distinction between self-exciting and doubly stochastic mechanisms, we have made that distortion explicit and constructed a delayed bivariate family whose projected kernels are nonnegative and stable. The admissibility calculation is essential: a causal spectral representation alone need not define a point process with nonnegative intensity.

The available information determines which structural conclusions are possible. Integrated third-order cumulants with Cox immigration identify normalized finite-dimensional parameters generically and locally in the stated dimensions, but not kernel shapes or the full factor law. For the symmetric pair, an exact finite-window factorial-cumulant contrast distinguishes mechanisms with the same mean and full spectrum. The controlled experiment demonstrates this separation and its deterioration under a weaker signal; it is not a universal test for excitation. Restricted Whittle estimation addresses a different problem: parametric restrictions can produce informative spectrum-based estimation without eliminating unrestricted second-order ambiguity. We therefore state spectral identified-set consistency separately from regular interior normality and mean-corrected feasible inference.

The financial-data illustration reinforces the need to state the fitted information set. Nearby within-hour criteria can coexist with sharply different implications for hourly totals and long-run excitation, and the omitted moments can reveal poor adequacy of the retained fits. This does not establish empirical equality of full second-order laws. The mathematical and numerical results together distinguish a predictive representation, an admissible Hawkes counterpart, and identification of the underlying mechanism; none should be substituted for another.

\appendix
\section{Proof of the delayed nonnegative counterpart}\label{app:delayMain}
This appendix proves Corollary~\ref{cor:exactdelay}. The filtering calculation supplies the four kernels; the row-sum calculation below separately verifies that they define a stationary Hawkes process with the required mean. These are distinct steps, because causality of a spectral factor alone does not imply Hawkes admissibility.

\begin{proof}
Use a Gaussian covariance surrogate for the centered common signal $X_t=\beta(Y_t-\E Y)$:
\[
\dd X_t=-\kappa X_t\dd t+\sqrt{2\kappa\beta^2v}\,\dd W_t,
\]
with mutually independent Brownian inputs $W,V_1,V_2$ and observations
\[
\dd O_1(t)=X_t\dd t+\sqrt\Lambda\,\dd V_1(t),
\qquad
\dd O_2(t)=X_{t-\Delta}\dd t+\sqrt\Lambda\,\dd V_2(t).
\]
All parameters $a,b,c,q,D,E,F_0$ are as in Corollary~\ref{cor:exactdelay}; in the intermediate calculation $D(s)$ denotes the same normalizing function at horizon $s$, so $D=D(\Delta)$. At calendar time zero, re-index the slow observation into latent time. There are two independent measurements of $X_s$ for $s\le-\Delta$ and only the fast observation for $-\Delta<s\le0$. The stationary filter with two measurements has error covariance $P_0=\Lambda q$ and decay rate $a$. Hence its estimate $m_0$ of $X_{-\Delta}$ uses weight $q e^{-a(-\Delta-s)}$ on each of the two latent-time measurement streams.

On the remaining interval of length $\Delta$, the one-measurement normalized error covariance $k=P/\Lambda$ solves
\[
k'=\kappa^2r-2\kappa k-k^2,\qquad k(0)=q.
\]
Writing
\[
D(s)=\cosh(bs)+\frac cb\sinh(bs)
\]
gives $k(s)=D'(s)/D(s)-\kappa$ and propagation factor $h(s)=1/D(s)$. The current-state filtered mean is
\[
m(s)=h(s)m_0+\int_0^s \frac{h(s)}{h(t)}k(t)\,\dd O_1(t-\Delta).
\]
At $s=\Delta$, the coefficient of $\dd O_1(t-\Delta)$ is
\[
\frac{h(\Delta)}{h(t)}k(t)
=\frac{D'(t)-\kappa D(t)}{D(\Delta)}
=\frac{qF_0(t)}{D(\Delta)}.
\]
The coefficient of $m_0$ is $1/D(\Delta)$. Converting $t=\Delta-u$ to calendar lag gives \eqref{eq:delay11}--\eqref{eq:delay12}, with the factor $q/D$ on every branch.
For the estimate of $X_{-\Delta}$, the conditional cross-covariance with the current state is $P_0h(s)$. The fixed-point smoother is therefore
\[
m_{\mathrm{sm}}=m_0+\int_0^\Delta qh(s)
 \{\dd O_1(s-\Delta)-m(s)\dd s\}.
\]
Substitute the formula for $m(s)$ and use
\[
\int_0^s h(t)^2\dd t=\frac{\sinh(bs)}{bD(s)}.
\]
The resulting coefficients are $E(\Delta)/D$ on $m_0$ and $qE(\Delta-t)/D$ on $\dd O_1(t-\Delta)$. This gives \eqref{eq:delay21}--\eqref{eq:delay22}.
All entries are nonnegative. Direct integration gives
\[
1-\bar\Psi_{11}-\bar\Psi_{12}
=\frac{\kappa}{D}\left\{\frac1a+\frac{\sinh(b\Delta)}b+\frac{c}{b^2}[\cosh(b\Delta)-1]\right\}>0,
\]
and
\[
1-\bar\Psi_{21}-\bar\Psi_{22}
=\frac{\kappa}{D}\left\{\frac{\kappa c}{ab^2}\cosh(b\Delta)+\frac{\kappa}{ab}\sinh(b\Delta)+\frac q{b^2}\right\}>0.
\]
Thus $\rho(\bar\Psi)<1$ and the Hawkes baseline $\Lambda(I-\bar\Psi)\1$ is positive.

For the monotonicity statement, integration of the cross-kernels gives
\[
\bar\Psi_{12}=\frac q{aD},
\qquad
\bar\Psi_{21}=\frac qD\left\{\frac{\sinh(b\Delta)}b+\frac\kappa{b^2}[\cosh(b\Delta)-1]+\frac{E(\Delta)}a\right\}.
\]
Their ratio reduces to \eqref{eq:Qratio}. Its power series is
\[
\begin{aligned}
Q(r,x)=1&+\sum_{n\ge1}\left[(1+r)^n+\sqrt{1+2r}(1+r)^{n-1}\right]\frac{x^{2n}}{(2n)!}\\
&+\sum_{n\ge0}\left(\sqrt{1+2r}+1\right)(1+r)^n\frac{x^{2n+1}}{(2n+1)!}.
\end{aligned}
\]
For $x>0$ every displayed coefficient is strictly increasing in $r$, so $Q$ and hence $(Q-1)/(Q+1)$ are strictly increasing. As $r\downarrow0$, $Q\to2e^x-1$, giving $1-e^{-x}$; as $r\to\infty$, $Q\to\infty$.
\end{proof}

\section{Exact rank certificates and realizability}\label{app:certificatesMain}
\begin{proof}[Proof of Theorem~\ref{thm:id}]
The entries of $M_\Lambda(\theta)$ are rational functions of $(\Lambda,\theta)$ on the stability region, with denominator a power of $\det(I-\Phi)$. Hence any maximal Jacobian minor is rational. A single nonzero rational witness proves that the rank-deficient set is a proper algebraic subset in the joint variables.

For $d=2$, take
\[
\Phi=\begin{pmatrix}1/5&1/10\\1/20&1/4\end{pmatrix},
\qquad
\beta=\begin{pmatrix}1/3\\1/2\end{pmatrix},
\qquad
\Lambda=\begin{pmatrix}2\\3\end{pmatrix},
\qquad
s_3=0.
\]
Using row-major $\Phi$, then $\beta_1,\beta_2,s_3$, and the ordered moments $(C_{11},C_{12},C_{22},K_{111},K_{112},K_{122},K_{222})$, exact rational differentiation gives
\[
\det J_2=-\frac{2^{37}5^{23}\cdot19\cdot47\cdot359}{3^2 7^{14}17^{14}}\ne0.
\]
For $d=3$, take
\[
\Phi=\begin{pmatrix}
1/5&1/10&1/20\\
1/20&1/4&1/10\\
1/10&1/20&1/6
\end{pmatrix},
\quad
\beta=\begin{pmatrix}1/3\\1/2\\2/5\end{pmatrix},
\quad
\Lambda=\begin{pmatrix}2\\3\\4\end{pmatrix},
\quad s_3=0.
\]
Using all six distinct entries of $C$ and $K_{111},K_{112},K_{113},K_{122},K_{123},K_{133},K_{222}$, the resulting $13\times13$ minor has determinant
\[
\det J_{3,I}=-\frac{2^{98}3^7 5^{35}7^2\cdot73\cdot84233309\cdot B}{13^{21}29^{22}31^{23}}\ne0,
\]
where
\[
B=1319764221011104608386215199.
\]
Both witnesses are realizable interior points. For example, choose a symmetric $\{0,1\}$ chain with total switching rate $\kappa=1/2$. Then $\sigma^2=1$ and $s_3=0$; with $\E Y=1/2$, the corresponding nonnegative baseline vectors are $(17/15,19/10)^\T$ and $(14/15,3/2,167/60)^\T$, respectively.

The witnesses lie inside a realizable moment-parameter family, not merely in an ambient rational domain. To see this, let a $\{0,1\}$ chain have stationary probability $p$ and total rate $\kappa=2p(1-p)$. Then $\sigma^2=1$ and
\[
s_3(p)=\frac{3(1-2p)}{2p(1-p)},\qquad s_3'(1/2)=-12\ne0.
\]
Thus $s_3$ varies on an open interval around zero while the strictly positive baselines at both witnesses remain positive under sufficiently small perturbations of $(\Lambda,\Phi,\beta,p)$. The nonzero minors imply full column rank off a proper algebraic set. Fubini's theorem then gives the asserted quantifier for almost every fixed positive $\Lambda$; a nonsingular square submap gives local identification by the inverse function theorem.

These witnesses establish the joint-generic rank in Theorem~\ref{thm:id}, not the stronger quantifier ``for every fixed positive $\Lambda$.'' No rank conclusion for larger dimensions follows from these two certificates.
\end{proof}

\section*{Supplementary Materials}
The Online Supplement contains the remaining proofs, regularity verification, full experimental protocols, and additional empirical diagnostics. The main paper includes the projection proof, the delayed-counterpart proof, the realizable exact identification certificates, and the finite-window separation proof.

\section*{Data Availability Statement}

The financial transaction data used in this study are
publicly available from the Binance Public Data repository
at \url{https://data.binance.vision/}.
The empirical analysis uses SOLUSDT and BTCUSDT spot and
USD-margined perpetual-futures transaction records for
August 2026. The simulation results are based on
synthetically generated data under the models and parameter
settings described in the manuscript and Online Supplement.

\section*{Declaration of Generative AI Use}
The authors used OpenAI ChatGPT (GPT-6) during manuscript preparation to assist with language refinement, exposition, organization, and coding. All AI-assisted material was reviewed and revised by the authors. The authors verified the mathematical statements, references, code, and reported numerical results, and take full responsibility for the accuracy, originality, and integrity of the manuscript.

\begingroup
\small\setlength{\bibsep}{3pt}

\endgroup

\end{document}